\documentclass[openany]{aa} 
\pdfoutput=1 
\usepackage[varg]{txfonts} 

\usepackage{graphicx}
\usepackage{txfonts}
\usepackage{rotating}
\DeclareGraphicsExtensions{.pdf, .png, .jpg, .jpeg, .tif, .mps, .bmp, .tiff, .gif}
\usepackage{amsmath}
\usepackage{amssymb}
\usepackage{color}
\usepackage[T1]{fontenc}
\usepackage{ae,aecompl}
\usepackage[normal]{subfigure}
\usepackage[font={small}]{caption}
\usepackage{natbib}

\usepackage{ulem}
\usepackage{placeins}
\usepackage{float}

\mathchardef\mhyphen="2D

\newcommand{\civ}{\ifmmode {\rm C}\,{\sc iv} \else C\,{\sc iv}\fi}
\newcommand{\CIV}{\ifmmode {\rm C}\,{\sc iv}\,\lambda1549 \else 
	           C\,{\sc iv}\,$\lambda1549$\fi}
\newcommand{\oiii}{O\,{\sc iii}}

\newcommand{\mgii}{Mg\,{\sc ii}}
\newcommand{\ciii}{C\,{\sc iii}}

\newcommand{\feii}{Fe\,{\sc ii}}
\newcommand{\feiii}{Fe\,{\sc iii}}

\newcommand{\siiv}{Si\,{\sc iv}}
\newcommand{\oiv}{O\,{\sc iv}}
\newcommand{\niv}{N\,{\sc iv}}
\newcommand{\heii}{He\,{\sc ii}}
\newcommand{\aliii}{Al\,{\sc iii}}

\newcommand{\alii}{Al\,{\sc ii}}
\newcommand{\siiii}{Si\,{\sc iii}}
\newcommand{\nii}{N\,{\sc ii}}
\newcommand{\hei}{He\,{\sc i}}

\begin{document}

   \title{Simultaneous detection and analysis of optical and ultraviolet broad emission lines in Quasars at $z\sim2.2$\thanks{Based on observations collected at the European Organisation for Astronomical Research in the Southern Hemisphere, Chile, under program 086.B-0320(A).}}


\author{S.~Bisogni\inst{\ref{inst1},\ref{inst2}}\thanks{E-mail: susanna@arcetri.astro.it}\and S.~di~Serego~Alighieri\inst{\ref{inst2}}\and
P.~Goldoni\inst{\ref{inst3}}\and L.~C.~Ho\inst{\ref{inst4},\ref{inst5}}\and A.~Marconi\inst{\ref{inst1},\ref{inst2}}\and
G.~Ponti\inst{\ref{inst6}}\and G.~Risaliti\inst{\ref{inst1},\ref{inst2}}}

\institute{Dipartimento di Fisica e Astronomia, Universit\`a di Firenze, via G. Sansone 1, 50019 Sesto Fiorentino, Firenze, Italy\label{inst1} 
\and
INAF - Osservatorio Astrofisico di Arcetri, Largo E. Fermi 5, I-50125, Firenze\label{inst2}
\and
APC, Astroparticules et Cosmologie, Universit\'e Paris Diderot, CNRS/IN2P3, CEA/Irfu, Observatoire de Paris, Sorbonne Paris Cit\'e,
10 rue Alice Domon et L\'eonie Duquet, 75205 Paris Cedex 13, France\label{inst3}
\and
Kavli Institute for Astronomy and Astrophysics, Peking University, Beijing 100871, China\label{inst4}
\and
Department of Astronomy, School of Physics, Peking University, Beijing 100871, China\label{inst5}
\and
Max-Planck Instit\"ut f\"ur extraterrestrische Physik, Giessenbachstrasse 1, 85748 Garching bei M\"unchen, Germany\label{inst6}
}

   \abstract{We studied the spectra of six $z \sim 2.2$ quasars obtained with the X-shooter spectrograph at the Very Large Telescope. The redshift of these sources and X-shooter's spectral coverage allow us to cover the rest spectral range $\sim1200 \mhyphen 7000$\AA~ for the simultaneous detection of optical and ultraviolet lines emitted by the Broad Line Region. Simultaneous measurements, avoiding issues related to quasars variability, help us understanding the connection between different Broad Line Region line profiles generally used as virial estimators of Black Holes masses in quasars.
The goal of this work is comparing the emission lines from the same object to check on the reliability of H$\alpha$, \mgii\ and \civ\ with respect to H$\beta$. H$\alpha$ and \mgii\ linewidths correlate well with H$\beta$, while \civ\ shows a poorer correlation, due to the presence of strong blueshifts and asymmetries in the profile.
We compare our sample with the only other two whose spectra were taken with the same instrument and for all examined lines our results are in agreement with the ones obtained with X-shooter at $z{\sim}1.5 \mhyphen 1.7$.
We finally evaluate \ciii] as a possible substitute of \civ\ in the same spectral range and find that its behaviour is more coherent with those of the other lines: we believe that, when a high quality spectrum such as the ones we present is available and a proper modelization with the \feii\ and \feiii\ emissions is performed, the use of this line is more appropriate than that of \civ\ if not corrected for the contamination by non-virialized components.}


\titlerunning{Simultaneous detection of optical and ultraviolet broad emission lines}
\authorrunning{S.~Bisogni et al.}

   \maketitle

\section{Introduction}
BH (Black Holes) are believed to be the origin of the extreme phenomenology of quasars and, more generally, of Active Galactic Nuclei (AGN) \citep{Rees1984}.
The growth of BHs occurs by accretion of the gas available in its proximity \citep{ShakuraSunyaev1973}.

The gravitational force and angular momentum of the SMBH forces the gas to rotate around it and the gas emission reveals the properties of the motion.
Broad (width > $2000 \, \textup{km s}^{-1}$) Emission Lines detected in the spectra of quasars are a consequence of this process. These lines are believed to form in the near proximity of the BH ($d<<1 \, \textup{pc}$) in the so-called Broad Line Region (BLR).
Given the spatial dimensions involved and the distances of these luminous sources, the study of the BLR emissions is one of the few ways to have an insight on the nuclear region, otherwise inaccessible to observations.
This is why the technique known as \emph{Reverberation Mapping} (or Echo Mapping) \citep{BlandfordMcKee1982, Peterson1993} has received so much attention.
Since its development, originally established with the purpose of a better knowledge of the geometrical properties of the BLR \citep{Bahcall1972}, the opportunity it gives to measure the mass of the central body has been realized \citep{Peterson2004,BentzKatz2015}.
As long as the gas in the central regions is forced to rotate under the gravitational influence exerted by the black hole, the width of the emission lines coming from these regions is a measurement of the BH mass, according to the equation
\begin{equation} \label{eq:1}
v^{2} = f \frac{G M_{BH}}{R_{BLR}}\;,
\end{equation}
where $f$, the virial factor, accounts for the geometry of the BLR \citep{Collin2006, NetzerMarziani2010}, still mostly unknown, R$_{BLR}$ is the dimension of the Broad Line Region as deduced by Reverberation Mapping, $v$ is the velocity of the emitting gas, and $G$ is the gravitational constant.
In addition, the presence of a relation between the size of BLR and the continuum luminosity \citep{Kaspi2000,Bentz2013} enables us to give a measurement of the central mass with a single spectroscopic observation in all the sources for which we can see the emissions coming from the BLR (type 1 AGN), avoiding the limits imposed by spatially resolved kynematic measurements and by the long observational times required by Reverberation Mapping.
The measurement of BH masses through the analysis of spectroscopic features then opens the possibility of a statistical study of these peculiar sources.

The BLR emission line most used for virial mass determination is by far H$\beta$ $\lambda4861$, mostly because this line is known to be emitted by gas in virialized conditions and also because of its prominence in the optical spectral window. For very distant sources the optical range is no more available and we have to find a replacement for H$\beta$. 
The most promising candidates for this role are \civ\ $\lambda 1549$ and \mgii\ $\lambda 2800$ \citep{McLureJarvis2002}, both in the UV rest spectral range of the sources; in using these lines we assume that they are emitted by approximatively the same region as H$\beta$, by gas in virialized conditions, and therefore that they have widths comparable with that of this line. However, it is well known that the BLR is stratified in terms of ionization potential \citep{DietrichKollatschny1993, PetersonWandel1999}. 
\civ\ has a much higher ionization potential than H$\beta$ and even more so than \mgii. Moreover, \civ\ emission has very different behaviors depending on the source and exhibits very often a blueshift and a general asymmetry. That does not fit well in an ordered keplerian motion scenario and it is believed to be associated with outflows or winds in the gas \citep{Murray1995, Richards2012}. 
It is then clear why the use of Balmer lines is more advisable in general and why \mgii\ can be considered a more reasonable replacement.  When the redshift of the source only allows the use of \civ, we are, however, compelled to find a different solution.
The most important point, therefore, consists in identifying which part of the line can be associated with the gas of the BLR in keplerian motion and which part we should instead consider in a non-virial state \citep{Denney2012}.
Furthermore, the emission variability of AGN is more important for shorter wavelengths and a comparison of \civ\ with the optical lines is not truly reliable if these lines are not simultaneously detected.
The simultaneous detection eluds the problems connected with the very fast variability of these lines, a typical signature of AGN spectroscopic emissions, and helps us in the search of possible connections between lines properties to obtain rules to be used when the optical virial estimator (H$\beta$) is not available.

\section{Sample selection, observations and data reduction}

\begin{table*} 
\small
\caption{Sample and observations properties. $z_{Shen}$ are the redshifts initially used to rest-frame the spectra \citep{Shen2011}, $z_{fit}$ is the redshift we estimate from the central wavelength of the [\oiii]$\lambda 5007$ line for every source, $D_{L}$ is the luminosity distance, $r$ is the SDSS r-band PSF magnitude and $E(B-V)$ the galactic extinction from \cite{Schlafly2011}. In determining the luminosity distance $D_{L}$ we use the following cosmological parameters: $H_{0}=70\, \textup{km s}^{-1} \textup{Mpc}^{-1}$, $\Omega_{M}=0.3$ and $\Omega_{\Lambda}=0.7$.}
\begin{center}
\renewcommand{\arraystretch}{1.2}
\begin{tabular}{l c c c c c c c c }
\hline
           Name                  & $z_{Shen}$ & $z_{fit}$ &  $D_{L}$    &    $r$       &   $E(B-V)$ &   Exp Time     & Airmass &  Seeing       \\
                                       &                  &                 &      [Gpc]      &    [mag]   &                   &          [s]         &                &     [$''$]       \\
\hline
J093147.37+021204.3   &$2.2867$&$2.29708\pm0.00016$&  $19.385$  & $18.69$ &$0.059$  & $1800$  &   $1.41$ &    $1.2$    \\
J103325.92+012836.3  &$2.1771$&$2.18428\pm0.00007$&  $17.307$  &$18.59$& $0.038$& $1800$     & $1.46$  &   $0.8$      \\
J105239.38-003707.3   &$2.2569$&$2.26327\pm0.00009$& $18.073$ &$18.31$&$0.049$& $2400$      &     $1.26$  &    $1.0$     \\
J121911.23-004345.5   &$2.2933$&$2.2976\pm0.0004$&  $18.407$  &$17.95$&$0.028$&  $2400$       &     $1.37$   &   $1.1$      \\
J123120.55+072552.6  &$2.3899$&$2.38369\pm0.00008$&  $19.254$  &$18.08$&$0.018$&   $3000$   &   $1.26$     &   $1.0$     \\
J124220.07+023257.6  &$2.2239$&$2.22036\pm0.00006$&  $17.656$  &$18.16$&$0.024$&   $1800$   &   $1.15$    &    $1.0$    \\
\hline
\end{tabular}
\label{tab1}
\end{center}
\end{table*}

X-shooter \citep{ver11} is a three-arm, single object echelle
spectrograph which started operations in October 2009.
The instrument covers simultaneously the wavelength range from 300 to 2400 nm
in the three arms: UVB ($\Delta \lambda = 300 - 550$ nm), VIS ($\Delta \lambda =
550 - 1020$ nm) and NIR ($\Delta \lambda = 1020 - 2400$ nm), respectively.
For our observations we used slit widths of 1.3, 1.2 and 1.2 arcsec
respectively for the three arms resulting in resolving powers $R = \lambda/\Delta
\lambda =$ 4000, 6700 and 4300.

 The sample was selected with the goal of extending the work presented in \citep[][, Paper I hereafter]{Ho2012} at 
higher redshift. In that paper a sample of relatively bright (r$\sim$ 18-19) quasars from
the SDSS DR7 release \citep{aba09} with redshift around $\sim$ 1.5 was analyzed. The
redshift choice ensures a simultaneous coverage from \civ\ to H$\alpha$ with X-shooter.
For this effort we selected, again from the SDSS DR7 release, QSOs
with redshift around $\sim$ 2.3 ensuring again that X-shooter would detect
\civ\ to H$\alpha$ shifted at higher wavelengths with respect to the
previous sample. In order to obtain higher S/N spectra, especially
in NIR where the spectra are noisier, we selected slightly brighter
(r$\sim$ 17.5-18.5) QSOs observable in a single night at the VLT.
The resulting sample contained eight QSOs.
After selection we also checked from the SDSS spectra that the selected objects have
broad emission lines suitable to BH mass estimation and that they have
no obvious broad absorption features. The average broad lines FWHM of
the sources in the sample is consistent with the average at these
redshifts. However the average bolometric luminosity is <logL$_{bol}$>=47.25, higher
than the average bolometric luminosity of QSOs at this redshift, log
L$_{bol}$=46.8 $\pm$ 0.3 \citep{Shen2011}, but compatible within 1.5 $\sigma$. This
ensures that our sample is not strongly biased.

 Observations were performed in the framework of the French Guaranteed
Time and took place on March the 10th 2011. For all our sources we report in Table
\ref{tab1} the properties and the characteristics of the
observations.

The night was not photometric and the observing conditions were changing.
Therefore during the night we monitored the spectra reduced on line
and we increased or decreased the observing time of the targets
depending on their quality. The night was also hampered by strong winds
whose speed was near (and sometimes over) the 12 m/s limit\footnote[1]{http://archive.eso.org/asm/ambient-server?site=paranal} which prevents pointing
towards Northern targets such as ours. 
These strong winds caused a loss of about two hours and  a half
of observing time on our program forcing us to drop two targets. The six observed
targets are listed in Table \ref{tab1} with exposure times, average airmass
and seeing.

 Each observation consisted of 4 different exposures of 450 sec to 750
sec each for a total of 1800 to 3000 sec.
The exposures were taken using the nodding along the slit technique
with an offset of 5 arcsec between exposures in a standard ABBA
sequence. The slit was put at parallactic angle. Every observation was
preceded by an observation of a telluric A0V standard at similar airmass.

 We processed the spectra using version 1.3.0 of the X-shooter
data reduction pipeline \citep{gol06,mod10}. The pipeline
performed the following actions. The raw frames were first subtracted 
and cosmic ray hits were detected and corrected using the
method developed by \cite{vdk01}. The frames were then divided by a master flat field
obtained by using day-time flat field exposures with halogen
lamps.The orders were extracted and
rectified in wavelength space using a wavelength solution
previously obtained from calibration frames.
The resulting rectified orders were then shifted and added
to superpose them thus obtaining the final 2D spectrum.
The orders were then merged and in the overlapping regions the
merging was weighted by the errors which were being propagated
during the process. From the resulting 2D merged spectrum
a one dimensional spectrum was extracted at the source's position.
The one dimensional spectrum with the corresponding error file
and bad pixel map is the final product of the reduction.

 To perform flux calibration we used different procedures
for the UVB data and for the VIS-NIR data.
In the UVB band we extracted a spectrum from
a staring observation of the flux standard LTT3218  \citep{ham92,ham94}
taken in the beginning of the night. We then reduced
the data using the same steps as above but in
this case we subtracted the sky emission lines using the
\cite{kel03} method. This spectrum was divided by the flux table of the same
star delivered with the pipeline to produce the response function. The response was then
applied to the spectrum of the sources.
For the VIS and NIR arm, we used the A0V stars as flux and
telluric standards. We extracted the A0V spectra with the same
procedure used for the flux standard. We then used these spectra to
apply telluric corrections and flux calibrations simultaneously
using the Spextool software \citep{vac03}. We then verified
if the final spectra of the three arms were compatible in the
common wavelength regions and performed a correction using the UVB
spectra as reference where needed. The spectral shapes
are compatible with the ones of the SDSS spectra, while the fluxes
are on average $\sim$ 50 $\%$ weaker.
Relying on the accurate SDSS flux calibration, we finally scaled our spectra
in order to match them with SDSS spectra in common wavelength regions reasonably free
from emission lines.


\begin{figure*}

   {\includegraphics[scale=0.165]{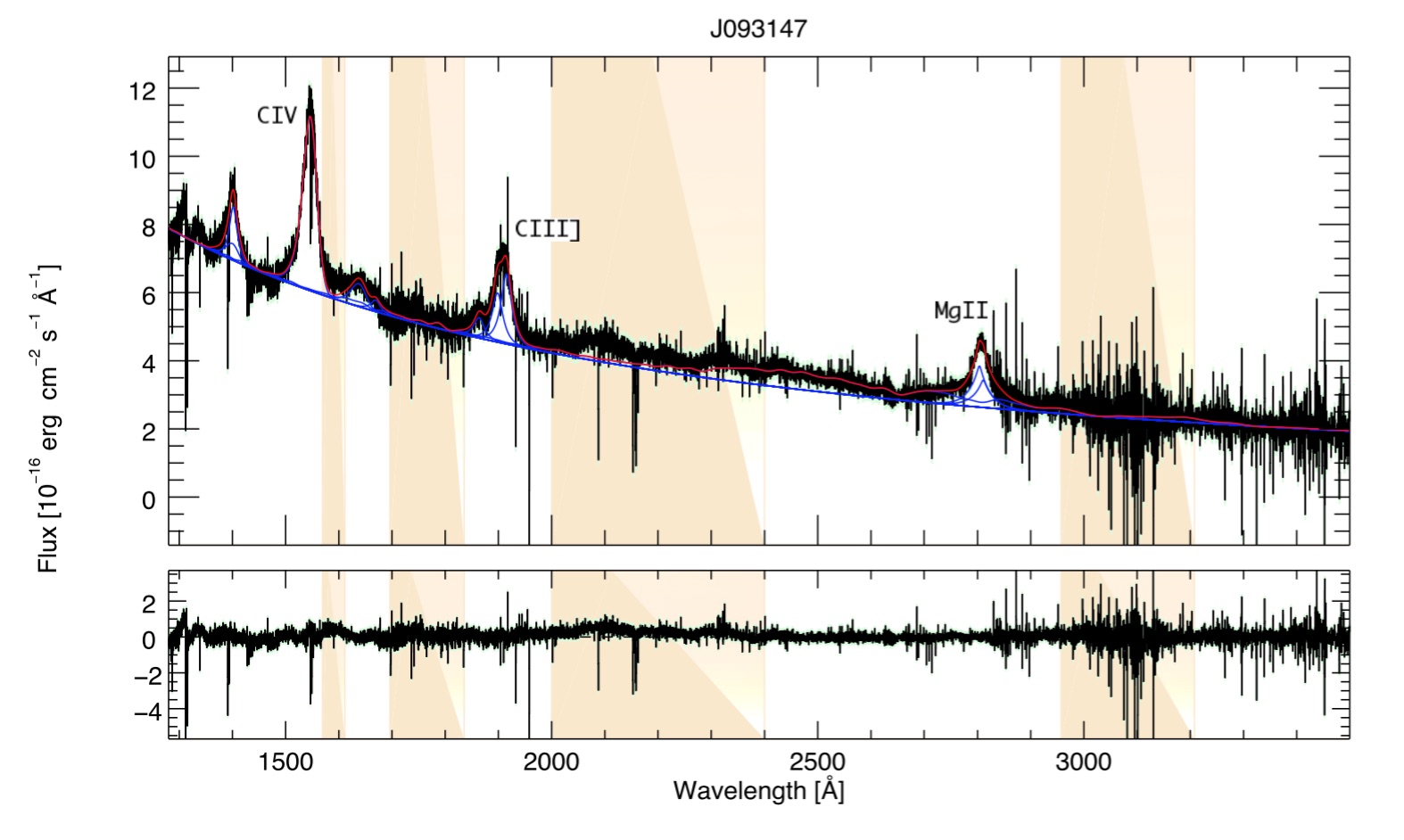}}   
   {\includegraphics[scale=0.165]{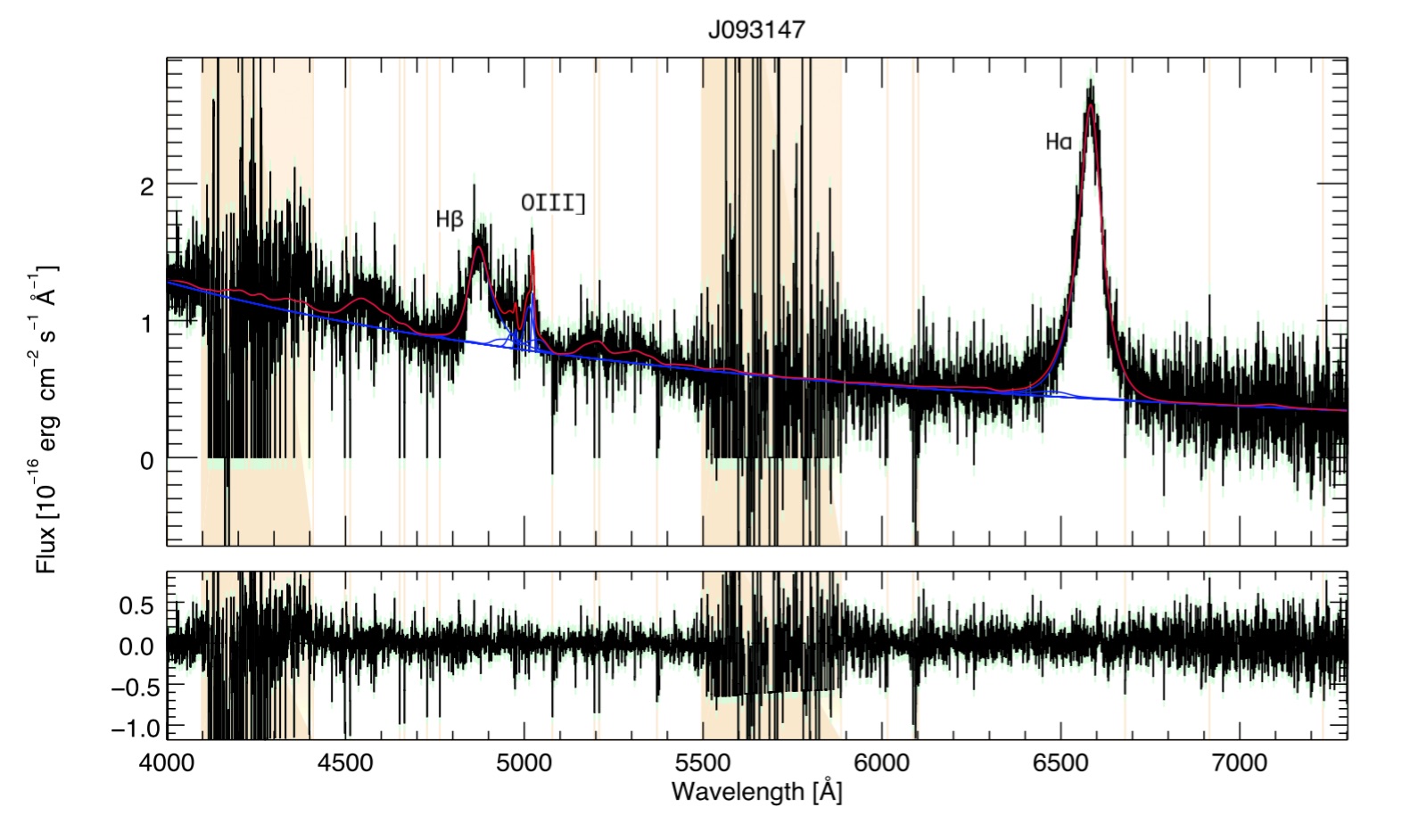}}
   {\includegraphics[scale=0.165]{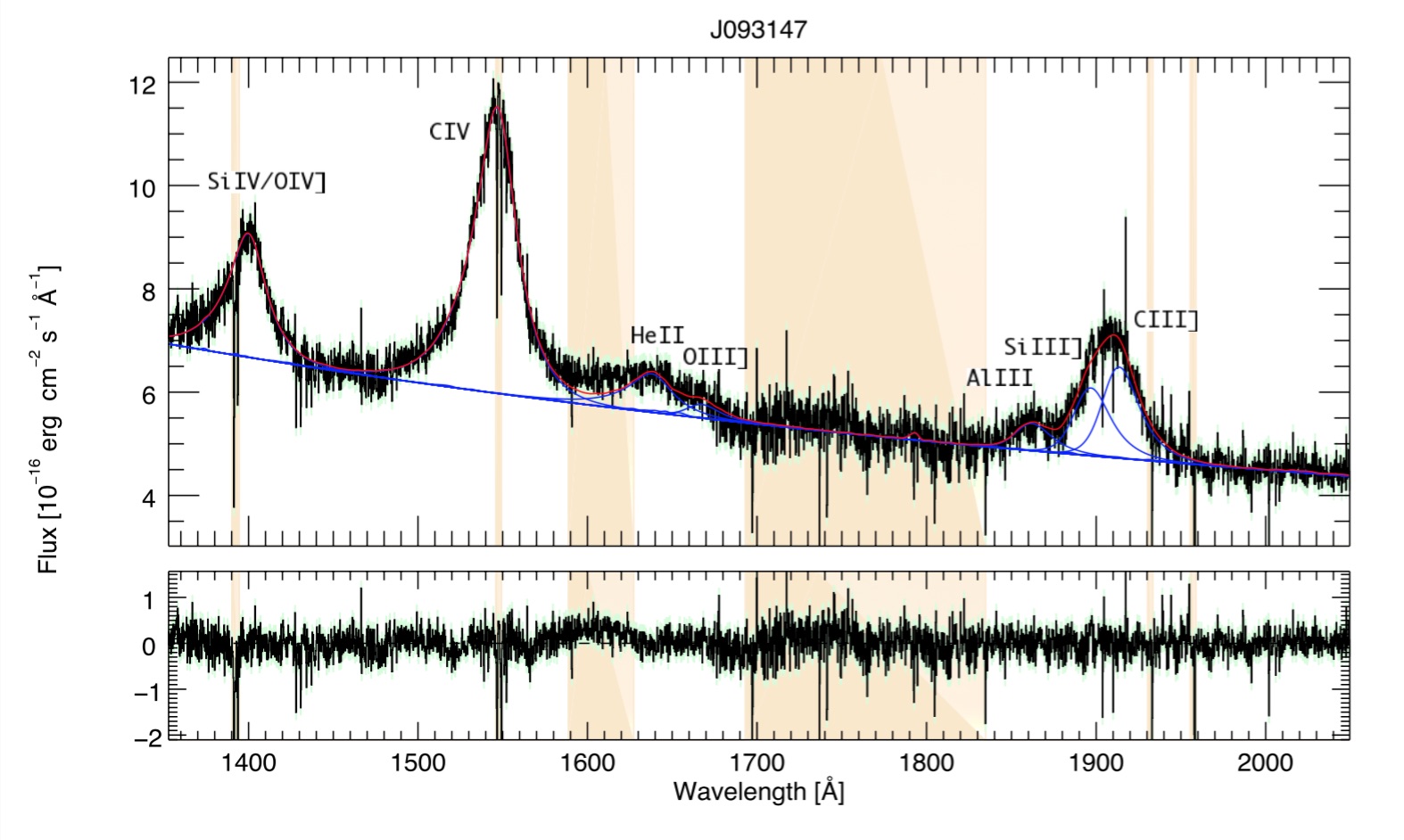}}
   {\includegraphics[scale=0.165]{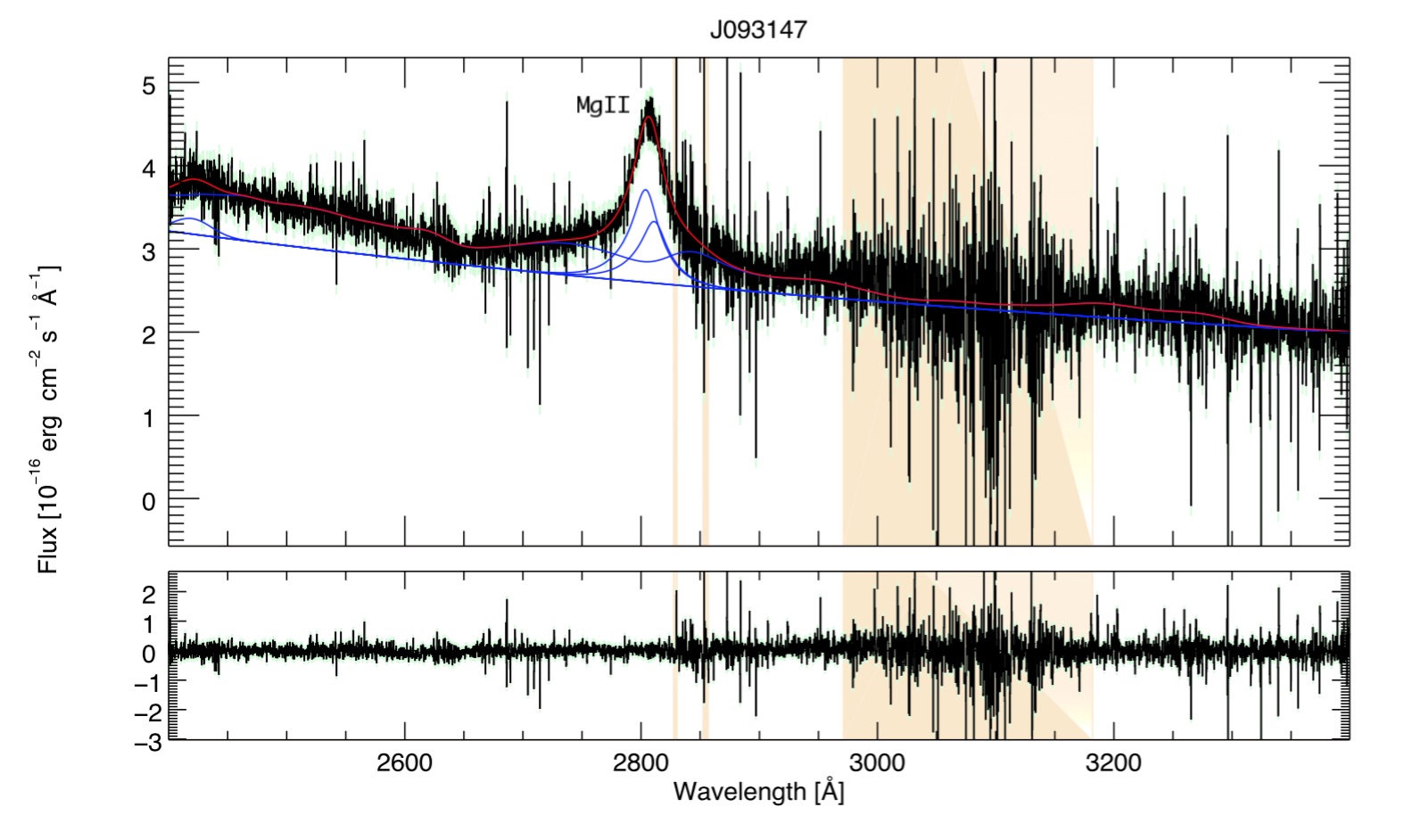}}
   {\includegraphics[scale=0.26]{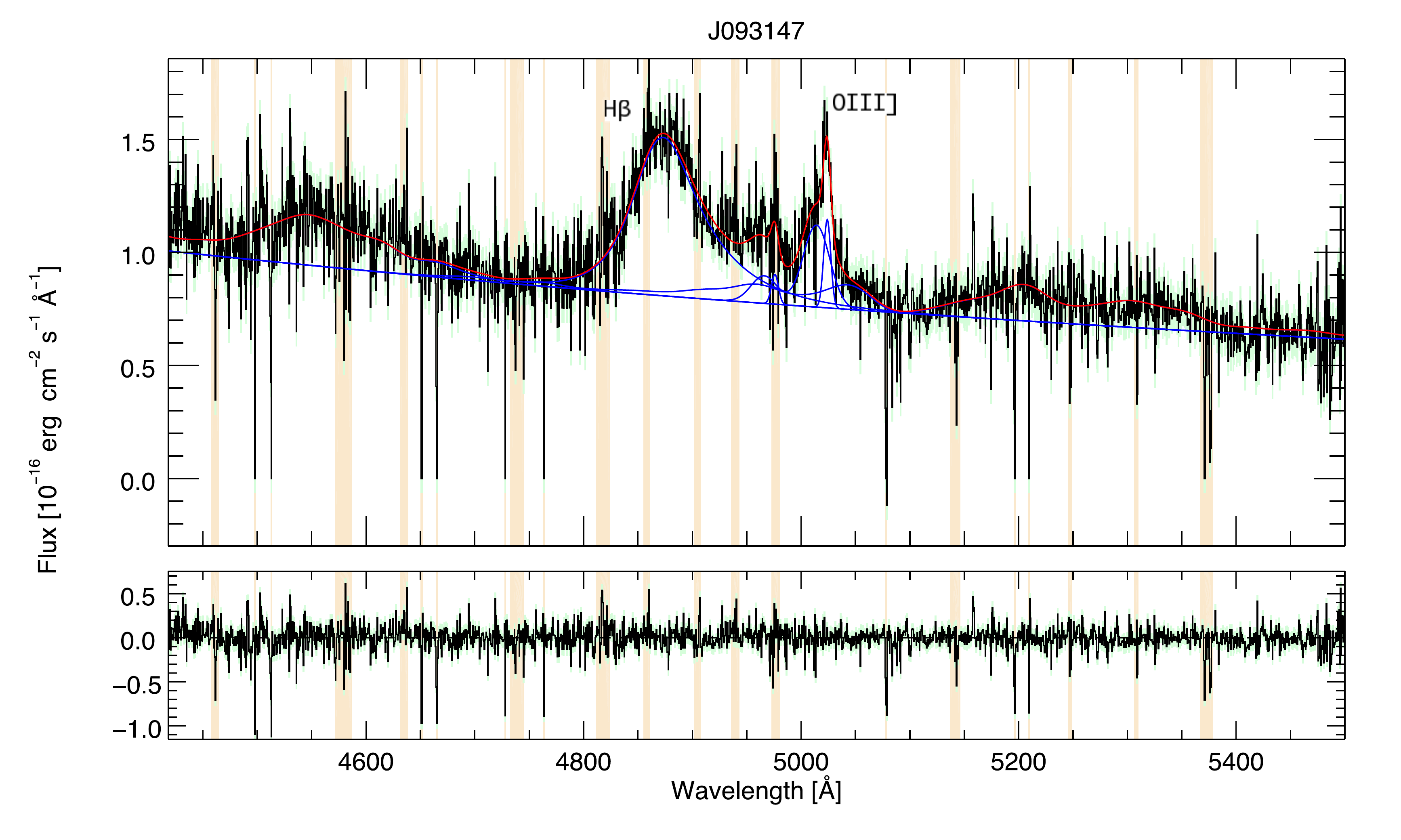}}
   {\includegraphics[scale=0.26]{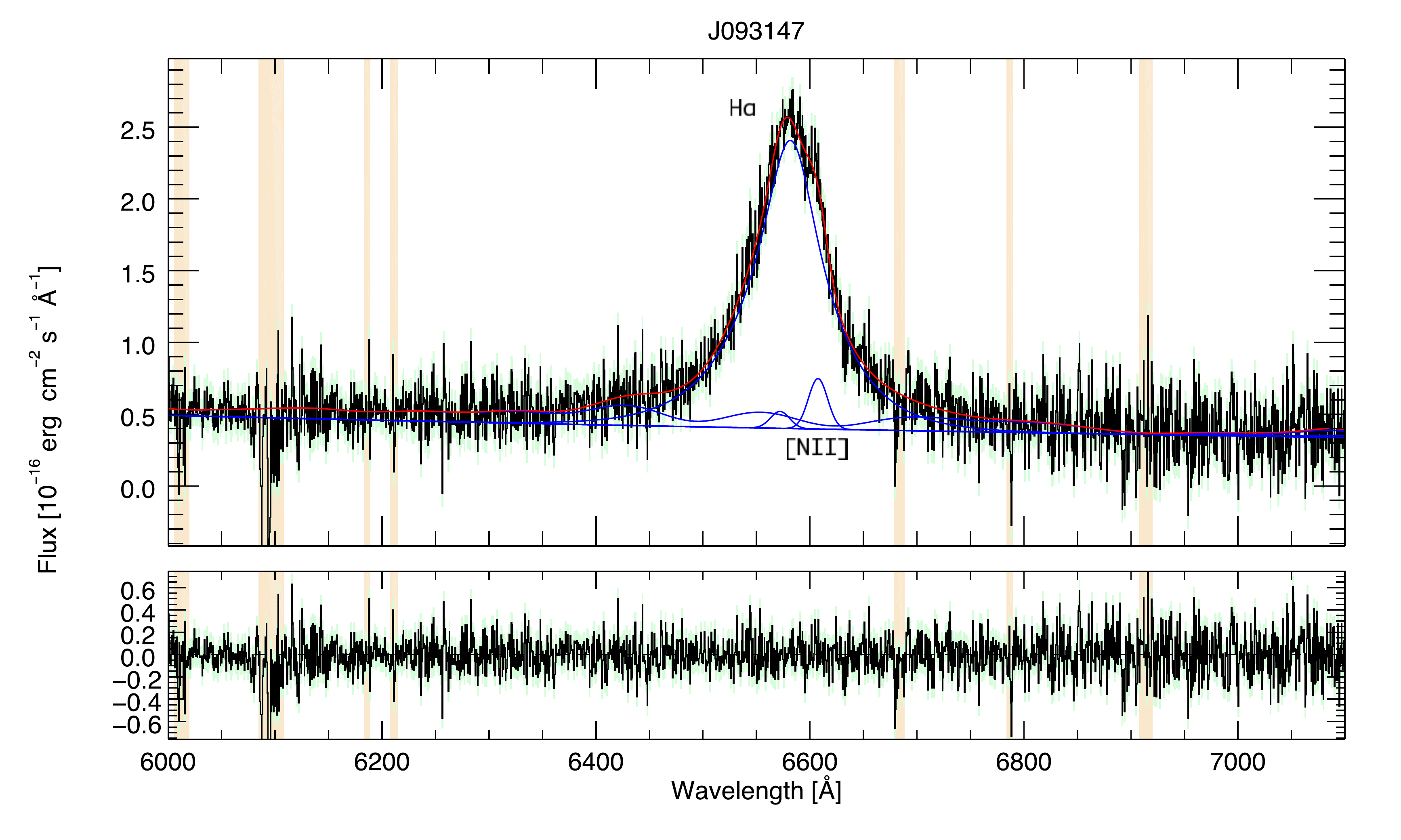}}
 \caption{Fits for all the examined spectral windows (\civ-\ciii]-\mgii\ and H$\beta$-H$\alpha$ large windows, \civ-\ciii], \mgii, H$\beta$ and H$\alpha$ small windows) for J093147. The black line is the original spectrum .The blue solid lines are the best fit models for the emissions (continuum, \feii\ and emission lines) and the red solid line is the total best fit. The lower panels show residuals between best fit model and original spectrum.
 The colored regions are those we chose to mask. This choice can be due to the presence of strong emission blending, to a non representativeness of the \feii\ templates or, in general, to a lack of knowlegde about what kind of emission is able to reproduce such features and the presence of noise.}
 \label{fig:fits_figuresJ093147}
\end{figure*}


\section{Spectral Fitting}
\label{sec:spectral_fitting}
As a preliminary step we de-redshifted the spectra according to their SDSS redshifts as reported in \cite{Shen2011} and corrected them for galactic extinction using the E(B-V) values from \cite{Schlafly2011} as listed in the NASA/IPAC Extragalactic Database\footnote[2]{The NASA/IPAC Extragalactic Database (NED) is operated by the Jet Propulsion Laboratory, California Institute of Technology, under contract with the National Aeronautics and Space Administration.} and the reddening law of \cite{Fitzpatrick1999} with R$_{V}=3.1$.

We then fitted the spectra with a procedure that uses the IDL MPFIT package \citep{Markwardt2009}, written with the purpose of a simultaneous fitting of continuum, \feii\ and other emission lines. 

Broad lines are fitted with a broken power law, convolved with a gaussian function to avoid the presence of a cusp at the peak \citep{Nagao2006};
the expression for the broken power law is
\begin{equation}
f(\lambda)\propto \begin{cases} \left(\frac{\lambda}{\lambda_{0}}\right)^{\beta} & \mbox{if}\; \lambda<\lambda_{0} \\ \left(\frac{\lambda}{\lambda_{0}}\right)^{\alpha} & \mbox{if} \; \lambda>\lambda_{0}  \end{cases}\;,
\label{eq:broaddoublepowerlawcomponent}
\end{equation}
where $\lambda_{0}$ is the central wavelength and $\alpha$ and $\beta$ are the slopes for red and blue tail respectively.
The choice of such a function allows us to reproduce with only five parameters (flux, $\lambda_{0}$, $\alpha$, $\beta$ and the $\sigma$ of the Gaussian function with which the double power law is convolved) the profiles which are commonly fitted with at least two Gaussian functions, involving $6$ parameters. This is particularly useful when dealing with emission line complexes, in which the use of a single component for every line helps in limiting the degeneracy in the fits. Moreover, when we do not fit lines separately, but several lines together, the use of a single fitting function allows us to set the same profile for all the lines with similar excitation conditions (high or low ionisation).
The results obtained by fitting with the function in Eq.~\ref{eq:broaddoublepowerlawcomponent} are consistent with those obtained by fitting with multiple Gaussians, as long as the total spectrum is well reproduced by the fit.
Narrow lines are instead fitted with a simple Gaussian, because their emissions are generally very well reproduced by this function. Where a blue asymmetry is present, as in the case of [\oiii] $\lambda5007$\AA, a second Gaussian takes into account this feature.

We first obtained the slope of the continuum for  fits for the entire UV (wavelength range $\sim 1400 \mhyphen 3500$\AA, containing \civ, \ciii] and \mgii) and for the optical window (wavelength range $\sim 4000 \mhyphen 7300$\AA, containing H$\beta$ and H$\alpha$).
We then used these slopes also in the fits for the four narrower windows, pertaining to \civ-\ciii], \mgii, H$\beta$ and H$\alpha$ emissions.
When necessary we applied a mask to the spectral regions contaminated by sky emissions (that was particularly required in the case of H$\beta$ spectral window).

For the UV spectral window we took into account several emission lines, following the prescriptions proposed in \cite{Nagao2006}: emissions are separated in two groups, high and low ionization lines (HIL: \oiv\ $\lambda1402.06$, \niv] $\lambda1486.496$, , \civ\ $\lambda1549.06$, \heii\ $\lambda1640.42$; LIL: \siiv\ $\lambda 1396.76$, \oiii] $\lambda1663.48$, \aliii\ $\lambda1857.40$, \siiii\ $\lambda1892.03$, \ciii] $\lambda1908.73$, \alii\ $\lambda2669.95$, \oiii\ $\lambda2672.04$), whose velocity profiles are known to have systematically different behaviours, so that some parameters pertaining to one group, such as the central velocity and the two power law indexes for blue and red tail, can be tied for each line.
We determined these fitting parameters using \civ\ for the HIL group and \ciii] for the LIL group.
The \mgii\ doublet was instead fitted independently of the other lines. This line should in theory be included in the LIL and therefore be tied to \ciii] parameters, but since \mgii\ is one of our investigation target it did not make sense to tie it to another line (considering also that the \ciii] complex, with three emission lines present, can be degenerate).
The optical window includes the Balmer lines, H$\beta$ and H$\alpha$, and a few other lines not always present in the spectra (H$\delta \, \lambda4103$, H$\gamma \, \lambda4342$, \hei\ $\lambda4472$, [\oiii] $\lambda4959,5007$, [\nii] $\lambda 6585,6550$ and \hei\ $\lambda7067$).  H$\beta$ and H$\alpha$ were fitted independently.

The fitting procedure also includes \feii\ emissions; they are reproduced convolving emission templates with a Gaussian that accounts for the velocity of the emitting gas.
We used two kinds of templates: the first one is the I Zw 1 \feii\ template by \cite{Veron2004}, valid only for the visible band, and the second one is a series of model templates obtained with the photoionization code Cloudy \citep{Cloudy2013}.

The Cloudy templates were computed with the following setup:
\begin{itemize}
\item we used the 371 levels \feii\ model \citep{Verner1999} instead of the simplified model \citep{Wills1985};
\item we considered a continuum emission similar to that examined in \cite{MathewsFerland1987}, resembling the spectrum of a typical radio-quite AGN;
\item we assumed a plane-parallel geometry with maximum cloud column density of $10^{23} \, \textup{cm}^{-2}$;
\item we used 10 combinations of ionizing photon flux and column density to consider the possible physical conditions of the BLR;
\item for all models we also considered the possibility of a $100 \, \textup{km/s}$ microturbulence velocity.
\end{itemize}
The above assumptions result in 20 different templates, which are shown in Tab. \ref{tab_feii_template}.

For the \civ-\ciii] spectral window, \feiii\ emissions were also taken into account. Specifically we used the \cite{VestergaardWilkes2001} empirical template for \feii\ and \feiii\ as deduced from the I Zwicky 1 spectrum. 

During the fitting procedure the templates are combined with a positive weight (free parameter of the fit) and convolved with a Gaussian function accounting for the velocity of the emitting gas (whose central value and $\sigma$ are free parameters of the fit as well).

\begin{table}
\small
\caption{Different combinations of the ionizing photons flux emitted by the primary source ($\Phi(H)$) and electronic density in the BLR clouds ($N_{e}$), used as input for Cloudy \feii\ simulated templates for different physical condition of the BLR. The two parameters define the space of possible values for the ionization parameter $U$, describing the physical condition in the ionized region. For every combination we considered the case a microturbulence velocity $u_{turb}=100$ km/s is present. The result is a set of 20 templates for the \feii\ emissions.}
\begin{center}
\def\arraystretch{2.0}
\begin{tabular}{l | p{0.25cm}p{0.25cm}p{0.25cm}p{0.25cm}p{0.25cm}p{0.25cm}p{0.25cm}p{0.25cm}p{0.25cm}p{0.25cm}}
\hline
log$\left(\frac{N_{e}}{cm^{-3}}\right)$                    &  $8$    & $10$   & $10$   &   $12$ &  $12$  &   $ 12$   &   $14$ &   $14$   &   $ 14$  &   $14$  \\
\hline 
log$\left(\frac{\Phi(H)}{cm^{-2}\;s^{-1}}\right)$       &  $17$   & $17$   & $19$   &   $17$ &  $19$  &   $ 21$   &   $17$ &   $19$   &   $ 21$  &   $23$  \\
\hline
\end{tabular}
\label{tab_feii_template}
\end{center}
\end{table}

\begin{table*}
\small
\caption{Measured quantities for the examined spectral windows.}
\begin{center}
\begin{tabular}{l c c c c c c c l }
\hline

                 & J093147      & J103325 &     J105239    &   J121911      &   J123120    &     J124220      \\

\hline
UV slope       &$-1.40\pm0.12$&$-1.40\pm0.17$& $-1.71\pm0.06$&$-1.87\pm0.05$&$-1.15\pm0.10$&$-1.47\pm0.18$\\
Optical  slope   &$-2.19\pm0.16$&$-2.15\pm0.25$&$-2.38\pm0.25$&$-2.44\pm0.15$&$-1.53\pm0.11$&$-1.76\pm0.17$\\
log($\lambda L_{\lambda}$)($1350$\AA)&$46.62\pm0.04$&$46.61\pm0.06$&$46.70\pm0.02$&$46.90\pm0.01$&$46.71\pm0.01$&$46.73\pm0.03$\\
log($\lambda L_{\lambda}$)($1450$\AA)&$46.62\pm0.04$&$46.59\pm0.06$&$46.69\pm0.01$&$46.87\pm0.01$&$46.71\pm0.01$&$46.71\pm0.03$\\
log($\lambda L_{\lambda}$)($3000$\AA)&$46.50\pm0.01$&$46.46\pm0.01$&$46.46\pm0.02$&$46.59\pm0.02$&$46.66\pm0.02$&$46.56\pm0.04$\\
log($\lambda L_{\lambda}$)($5100$\AA)&$46.22\pm0.02$&$46.21\pm0.04$&$46.14\pm0.04$&$46.25\pm0.02$& $46.70\pm0.01$& $46.26\pm0.01$ \\
$v_{\textup{\civ}}$     &$-1487\pm21$&$-652\pm14$&$-1104\pm18$&$-916\pm15$&$21\pm21$&$-246\pm11$\\
FWHM$_{\textup{\civ}}$  &$5761\pm79$&$4049\pm52$&$5123\pm55$&$4804\pm48$&$5413\pm93$&$3846\pm19$\\
$\sigma_{\textup{\civ}}$   &$4160\pm63$&$3801\pm30$&$3792\pm56$&$3987\pm50$&$4690\pm48$&$3608\pm14$ \\ \relax
$v_{\textup{\ciii]}}$     &$-134\pm88$& $-254\pm31$&$330\pm26$&$-24\pm28$&$637\pm13$&$-29\pm22$ \\ \relax
FWHM$_{\textup{\ciii]}}$     &$4028\pm188$&$3357\pm108$ &$3368\pm75$&$5382\pm323$&$3557\pm45$&$3604\pm60$ \\ \relax
$\sigma_{\textup{\ciii]}}$     &$2328\pm203$&$2864\pm78$&$3085\pm75$&$3153\pm190$&$2603\pm45$&$2550\pm47$ \\ \relax
$v_{\textup{\mgii}}$     &$-167\pm31$&$54\pm24$&$123\pm18$&$567\pm34$&$165\pm19$&$53\pm15$\\
FWHM$_{\textup{\mgii}}$     &$3184\pm142$& $2294\pm126$&$2326\pm79$&$3345\pm82$&$3891\pm89$&$2680\pm71$ \\
$\sigma_{\textup{\mgii}}$     &$2466\pm79$&$2120\pm50$&$1592\pm45$&$2111\pm62$&$3641\pm96$&$2272\pm37$ \\
$v_{H_{\beta}}$     &$-239\pm60$& $-109\pm33$&$242\pm28$&$290\pm35$&$388\pm31$&$101\pm23$\\
FWHM$_{H_{\beta}}$     &$4483\pm174$&$2672\pm158$&$3093\pm107$ &$3771\pm127$&$3574\pm84$&$3378\pm111$\\
$\sigma_{H_{\beta}}$     &$3040\pm198$& $2542\pm121$&$2535\pm94$&$3261\pm122$&$3604\pm83$&$3087\pm73$\\
$v_{H_{\alpha}}$     &$-185\pm61$& $21\pm11$&$168\pm9$&$509\pm18$&$139\pm13$&$172\pm9$\\
FWHM$_{H_{\alpha}}$     &$3411\pm111$& $2638\pm42$ &$2246\pm30$&$3346\pm64$&$2497\pm175$ &$2537\pm32$\\
$\sigma_{H_{\alpha}}$     &$2608\pm161$&$2168\pm34$&$1769\pm25$&$2167\pm91$& $2555\pm183$ &$1982\pm26$\\
\hline
\end{tabular}
\label{tab3}
\end{center}
\end{table*}

None of the examined lines exhibits an evident narrow component. 
We decided then to fit all of the permitted lines with a single double power law function. 
Also the [NII] doublet in the H$\alpha$ window is not recognisable at all and we did not consider it among the emission lines in the fitting procedure.
Only for H$\alpha$ in J093147 the shape of the line profile reveals the presence of [NII] and, as a consequence of this, we considered the doublet in the fitting process.
A special case is represented instead by J123120, for which we originally considered only broad lines but the fit improved considerably taking into account an emission from the NLR too (see Section \ref{section:results} and Fig.~\ref{fig:fits_figures_J123120}). 
In Fig. \ref{fig:fits_figuresJ093147} an example for the UV and optical windows for one of the sources is presented, along with the individual windows for \civ, \ciii] and \mgii, H$\beta$ and H$\alpha$ lines (fits for the rest of the sources are shown in Appendix A). 
In the large UV window some wide regions were masked during the fitting process for the following reasons:
\begin{itemize}
\item[-] the presence of strong emission blending,
\item[-] a non representativeness of the \feii\ templates,
\item[-] a lack of knowlegde about what kind of emission is able to reproduce such features (this is for example the case of the red-shelf of \civ),
\item[-] the presence of noise.
\end{itemize}
The regions usually excluded from the fit are the red shelf of \civ, the bump of emission between \oiii] $\lambda1663.48$\AA~ and the \ciii] complex, when present, and the wide spectral region between $2000$ and $\sim 2450$\AA~ \citep[see][]{Nagao2006}.
This choice does not affect neither the continuum slope determination nor the line analysis, since the proper examination of the lines is performed on the individual spectral line windows. \\
For all the sources the FWHM and the $\sigma$ were estimated on the best fit profile for each line.
They are connected to the shape of the line; while the FWHM is more representative of the \emph{core}, the $\sigma$ depends more on the tails of the line \citep{Shen2013}. Choosing one or the other leads to different results in BH virial mass estimations, especially if we are dealing with poor quality data \citep{Denney2016}. \\
To give an estimate of the errors of these quantities we used a Monte Carlo approach, extracting $1000$ independent values for every parameter pertaining to the line.
On these synthetic profiles we evaluated $1000$ values for FWHM and $\sigma$; from the distributions of the $1000$ value of the FWHM and the $\sigma$ we were able to infer the relative error to be associated with a given measurement. As a consistency check, we also performed the fit of $100$ mock realizations of the original spectra, obtained with a random extraction, within its $1 \sigma$ uncertainty, of the flux in each channel. We measured the properties of the profile for all the 100 best fit and determined the standard deviation of the distribution. This is then the error associated with the property for the best fit of the true spectrum. This check was repeated for the lines in every spectral window. The errors on the measured quantities obtained in this way are consistent with those determined with the Monte Carlo approach.
The errors on the UV and optical slopes and on the luminosities at $1350$, $1450$, $3000$ and $5000$\AA\ are instead computed performing, for every source, fits of various regions of the spectra and then evaluating the differences between these results and those obtained in the original fit \citep{Pita2014}. All the measured quantities are listed in Tab. \ref{tab3}.  \\
We noticed that the central wavelength of the [\oiii] $\lambda5007$\AA~ line, from which the redshift was estimated \citep{Shen2011}, almost in every case was not close to the nominal wavelength. We then corrected the redshift using only the principal component of [\oiii] (excluding the blue shifted component from the whole profile, which is instead considered in the estimate of \citealt{Shen2011}). A more accurate estimate of the redshift of the sources can improve the analysis of the shifts of the lines with respect to their nominal wavelengths.
The corrected redshifts are listed in table Tab. \ref{tab1} and have been used in the following analysis.

\section{Results}
\label{section:results}
\subsection{Line comparison}
\begin{figure*}[h!]
\centering
\includegraphics[scale=0.80]{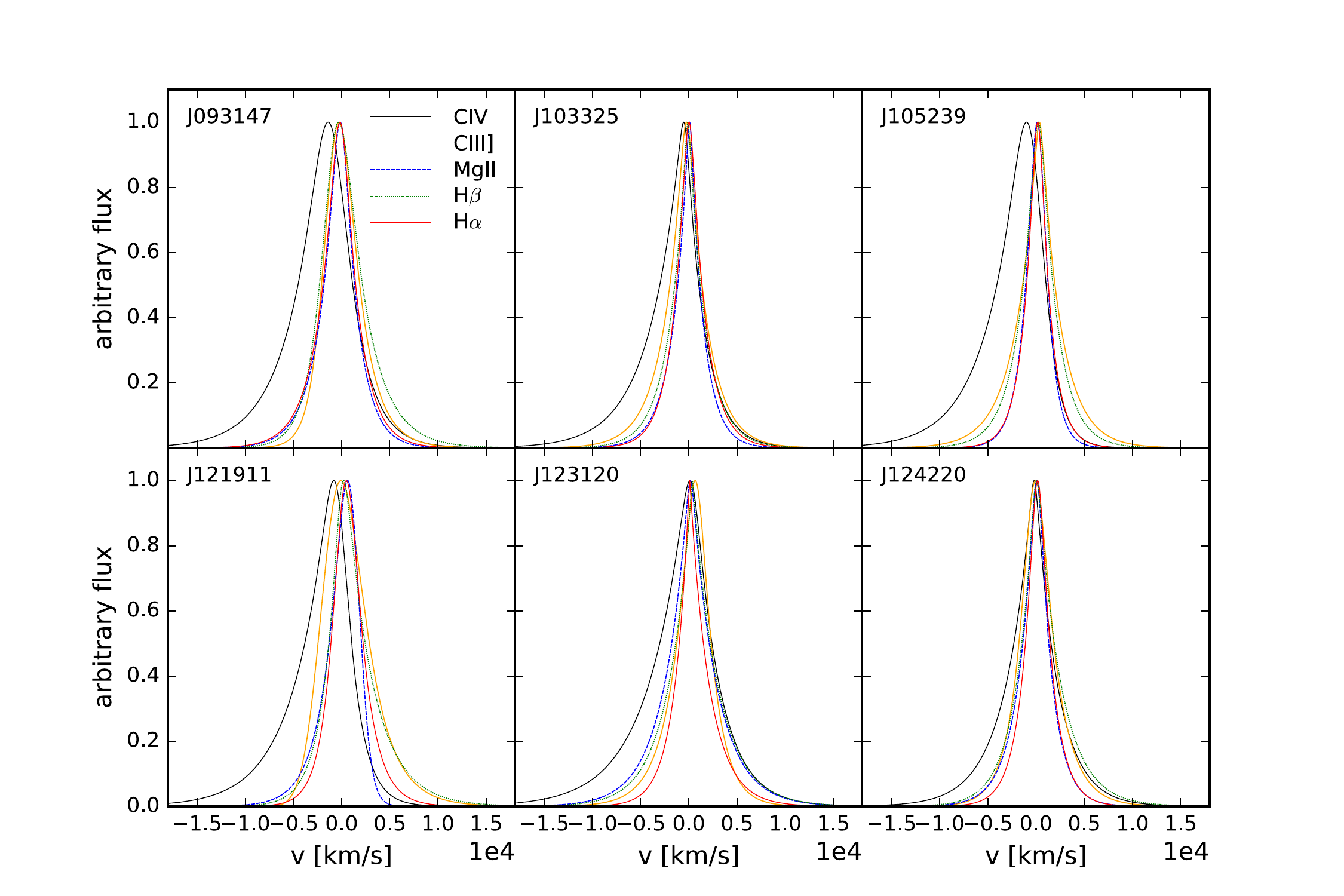}
\caption{Comparison of all the examined line profiles. The profiles are normalized to their peak values and referred to their laboratory wavelength on a velocity scale.}
\label{all_lines_comparison}
\end{figure*}

For a visual comparison we show in Fig. \ref{all_lines_comparison} the best fit profiles for every line in all spectra. The profiles are normalized to their peak values and presented on a velocity scale. 
As a general trend, H$\alpha$, H$\beta$ and \mgii\ behave similarly, according to what expected if the three lines are all emitted from regions in a virialized condition \citep{McLureJarvis2002,GreeneHo2005,Marziani2013}.
All of them show symmetric profiles and small shifts in the central wavelengths, generally below $300$km/s (Tab. \ref{tab3}).
Surprisingly, the most asymmetric line among these (usually in the red wing) is H$\beta$.
We suspect this is the result of a possible degeneration within the H$\beta$-[\oiii] complex, especially when \hei\ and \feii, whose emission are difficult to disentangle by the fitting procedure, are present.
The most asymmetric line of all is \civ. 
This line frequently shows a significant shift in the central wavelength, about $-730 \, \textup{km s}^{-1}$ on average, and all cases show the presence of a prominent blueshift.
We notice that, in contrast with what found in Paper I, \ciii] does not seem to behave so differently with respect to the other lines.
All sources do not present a large shift (${\sim} 90 \, \textup{km s}^{-1}$ on average). 
Furthermore, in all sources \ciii] does not show a prominent blueshift as does \civ\ instead.

\subsection{Line widths comparison}

\begin{figure*}[hp!]
\centering
\includegraphics[scale=0.95]{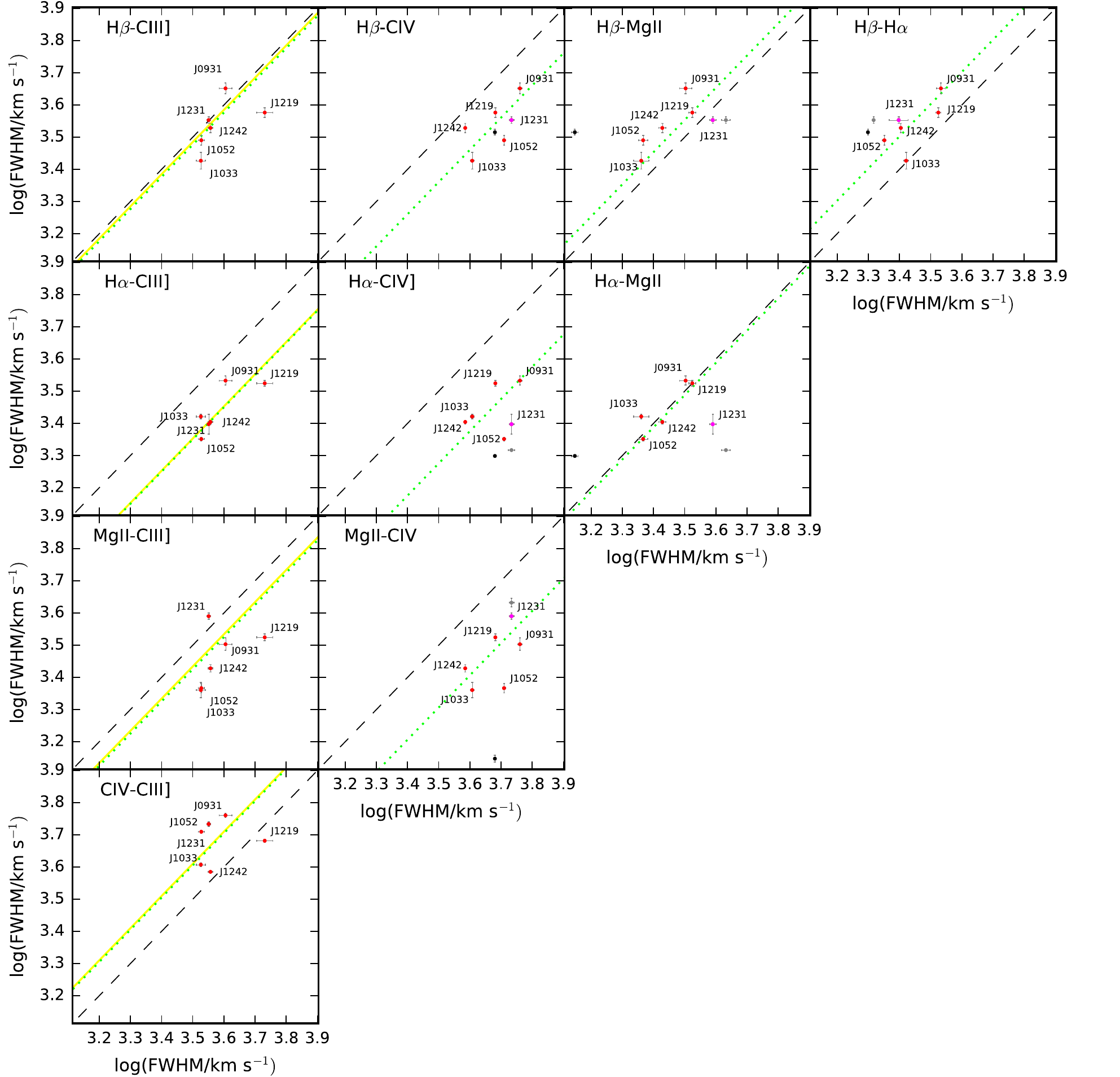}
\caption{FWHM correlations between line pairs. The lines labeled in the upper left of every panel refer to $Y$ and $X$ axis respectively. \\
$1^{st}$ column: correlations for the \ciii] line. The black dashed line represents the 1:1 relation, while the light green one represents the best fit to our data, considering a linear relation between the FWHMs for every pair of lines. The function we fit is log($\textup{FWHM}_{1}$)$ = m \cdot $log($\textup{FWHM}_{2}$)$ + c$, with $m=1$. The solid yellow line is the best fit when not considering J121911 in the sample. Intercept and scatter for all the relations are given in Tab.~\ref{tab6}. In the case of relations involving \ciii], we do not plot the different measurements for linewidths of J123120 (i.e. broad lines only vs broad and narrow lines considered in the fit). As in the case of the other lines, however, we take into account the FWHM measurements pertaining to the broad component when the narrow component is present.\\
$2^{nd}$, $3^{rd}$ and $4^{th}$ columns: correlations for all the lines commonly considered in virial estimates.
The red and magenta points are the measurements used in the final analysis. The black point represents the measurement taking into account only broad components for J123120, while the magenta point is the measurement including also narrow components in the fit. The grey point is an alternative measurement that still considers narrow emissions: in the case of \mgii\ it gives the FWHM of the line for which the width of narrow component was left as a free parameter of the fit (for the magenta point, instead, it was tied to that of \ciii]), while in the case of H$\alpha$ it gives the FWHM of the line for which the width of the narrow component was tied to that of [\oiii] and H$\beta$ narrow component (while for the magenta point was instead left free).
The black dashed line represents the 1:1 relation, while the light green one represents the bestfit to our data (red and magenta points), considering a linear relation between the FWHMs for every pair of lines. The function we fit is log($\textup{FWHM}_{1}$)$ = m \cdot $log($\textup{FWHM}_{2}$)$ + c$, with $m=1$. Intercept and scatter for all the relations are reported in Tab. \ref{tab4}.}
\label{fig:FWHM_comparison}
\end{figure*}
\begin{figure*}[hp!]
\centering
\includegraphics[scale=0.95]{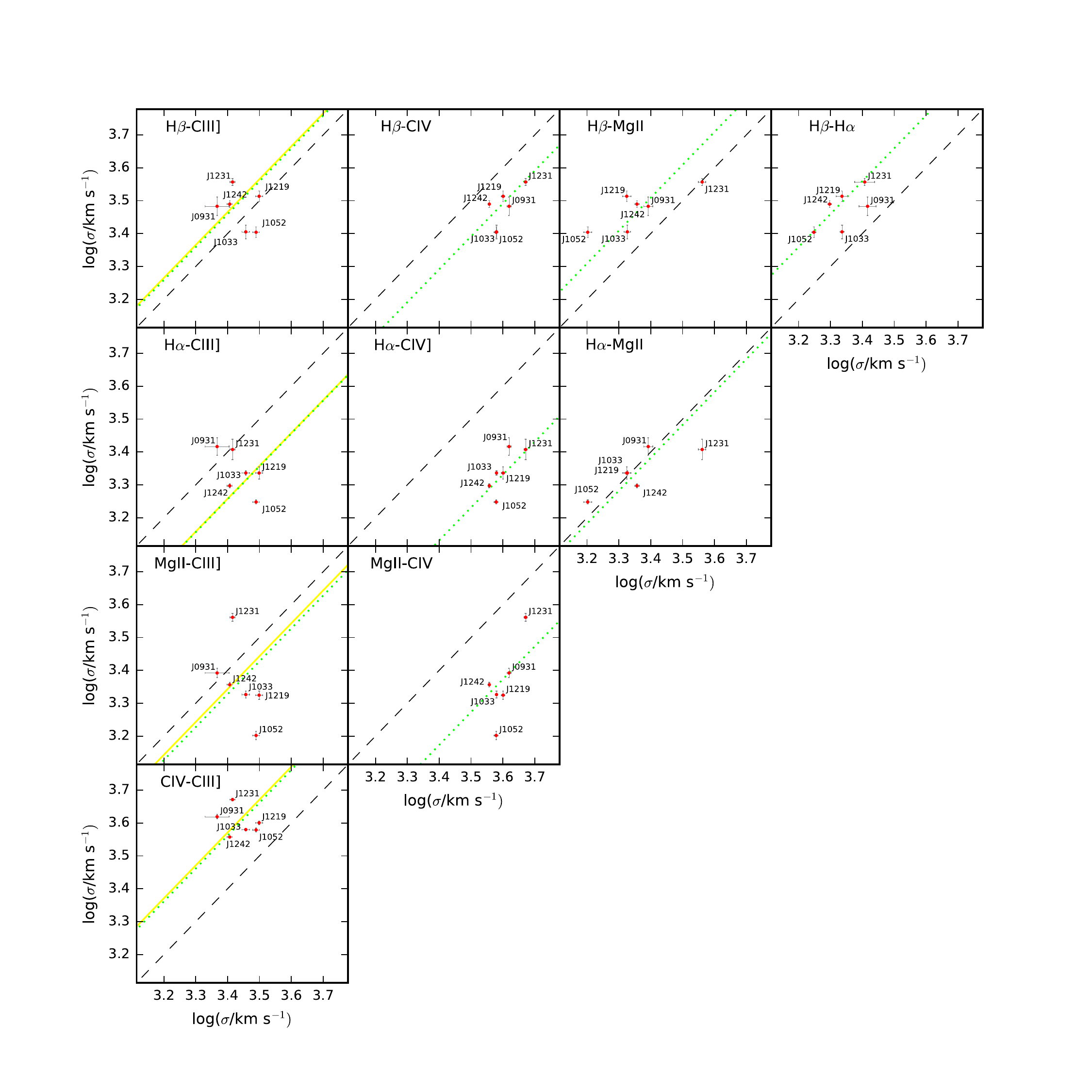}
\caption{Line dispersion correlations between line pairs. The lines labeled in the upper left of every panel refer to $Y$ and $X$ axis respectively. \\
$1^{st}$ column: correlations for the \ciii] line. The black dashed line represents the 1:1 relation, while the light green one represents the best fit to our data, considering a linear relation between the line dispersions. The function we fit is log($\sigma_{1}$)$ = m \cdot $log($\sigma_{2}$)$ + c$, with $m=1$. The solid yellow line is the best fit when not considering J121911 in the sample. Intercept and scatter for all the relations are given in Tab.~\ref{tab6}. \\
$2^{nd}$, $3^{rd}$ and $4^{th}$ columns: correlations for all the lines commonly considered in virial estimation. The black dashed line represents the 1:1 relation, while the light green one represents the bestfit for our data (red points), considering a linear relation between the line dispersions for every pair of lines. The function we fit is log($\sigma_{1}$)$ = m \cdot $log($\sigma_{2}$)$ + c$, where the slope $m=1$. Intercept and scatter for all the relations are reported in Tab. \ref{tab4}.}
\label{fig:sigma_comparison}
\end{figure*}


Fig. \ref{fig:FWHM_comparison} (second, third and fourth columns) shows a comparison  of FWHM for every pair of lines commonly examined for virial estimates, H$\beta$, \mgii, H$\alpha$ and \civ. The red points represent the measurements used in the final analysis. 

For J123120 we decided to do two fits, one using only the broad components (black points) and one including also the narrow ones.
The latter fit was done in two ways: the first one leaving the width of the narrow component as a free parameter and the other one fixing it to the the one of \ciii] in the UV range and to the one of [\oiii] and H$\beta$ in the optical range.
The magenta points were adopted for the following analysis and correspond to the case in which we tie the narrow component in the UV spectrum and to the case in which we left it free in the optical range.
While in the UV we could use the obvious technique of linking together the widths of the narrow components, we could not do the same in the optical because the narrow H$\alpha$ component is much broader than the [\oiii] one and they cannot be reasonably linked together.
The presence of a narrow component in J123120 is particularly evident for the \mgii\ line, although it improved the fits also fot the other lines.

Even for our small sample a correlation between FWHM of H$\beta$, H$\alpha$ and \mgii\ is present. 
Instead, \civ\ has a less strong correlation with the other lines: this is not surprising given the blueshift of the line in almost every source of the sample.
We report the results of the linear fit assuming a linear relation between the logarithms of the linewidths in Tab. \ref{tab4}.

Concerning the parametrization of the line width, although the $\sigma$ is in principle a more reliable estimator, especially when data quality is poor \citep{Collin2006, Denney2016}, most of the recent works use instead the FWHM (\citealt{Ho2012, TrakhtenbrotNetzer2012, HoKim2015} among the others), justifying it with a smaller scatter between different lines \citep{Mejia-Restrepo2016}.

In our measurements we do not observe such a larger scatter in the relationships involving the line dispersion $\sigma$ with respect to the FWHM (see Fig. \ref{fig:sigma_comparison} and Tab. \ref{tab4}). The only exceptions are the relationships involving \mgii, for which essentially the J123120 point is an outlier. Of course the smallness of our sample plays an important role in this respect, stressing the presence of outliers that would not probably be such in a larger sample. The analysis presented by \cite{Mejia-Restrepo2016} highlights the difference between measurements performed under a \emph{global} (considering a continuum fitted on the whole SED of accretion disk, BLR and NLR emissions) and a \emph{local} approach (the more common case, in which the fit on the line is performed only on a smaller spectral window including the line). The line measurements they report are those obtained under the local approach and for which they recognize the presence of a large scatter for the line dispersion, ascribable to the subtraction of a non proper fitted continuum. We notice, however, that their local approach take into account rather narrow spectral windows, while our measurements are performed on wider wavelength ranges and, moreover, take into account a preliminar continuum evaluation, performed on even wider windows.
Given the data quality and the spectral range that our fits cover, we are confident that our line dispersions could be considered for a virial estimate.
Nontheless, since we are especially interested in the comparison with some of the works mentioned before and our data quality allows the use of the FWHM, we will focus our analysis on this quantity.

\begin{table} 
\small
\caption{Linewidth correlations between line pairs. For both FWHM and line dispersion $\sigma$ we list the results for the linear fit between the logarithms: assuming a linear relation between the quantities, log($linewidth_{1}$)$= m \cdot $log($linewidth_{2}$)$ + c$, we find the intercept $c$ and the scatter $\Delta$ considering a fixed slope $m=1$. }
\begin{center}
\renewcommand{\arraystretch}{1.2}
\begin{tabular}{l c c c c c l }
    \hline
    & \multicolumn{2}{c}{log(FWHM)} &  \multicolumn{2}{c}{log($\sigma$)}\\

                                      &         $c$          &      $\Delta$    &         $c$        &    $\Delta$        \\
\hline
H$\beta$-\ciii]                &    $-0.024$         &       $0.046$   &    $0.059$     &    $0.081$            \\
H$\beta$-\civ\               &       $-0.139$   &    $0.058$    &    $-0.109$        &        $0.039$            \\
H$\beta$-\mgii\               &       $0.053$   &    $0.068$    &    $0.109$        &        $0.073$            \\
H$\beta$-H$\alpha$     &     $0.103$      &    $0.045$         &    $0.159$        &    $0.044$              \\
H$\alpha$-\ciii]              &    $-0.150$      &   $0.032$        &   $-0.144$    &   $0.066$              \\
H$\alpha$-\civ\             &    $-0.224$      &   $0.078$        &   $-0.269$    &   $0.033$              \\
H$\alpha$-\mgii\             &    $-0.012$     &     $0.048$    &   $-0.018$    &       $0.049$                \\
\mgii -\ciii]                      &    $-0.075$     &     $0.093$         &   $-0.072$      &   $0.137$              \\
\mgii-\civ\                       &   $-0.194$      &    $0.070$         &   $0.227$      &     $0.070$            \\
\civ-\ciii]                        &    $0.105$     &      $0.076$        &    $0.162$    &   $0.063$                  \\
\hline
\end{tabular}
\label{tab4}
\end{center}
\end{table} 

\FloatBarrier

\subsection{M$_{BH}$ and Eddington ratios}
Although we took into account several previous works \citep{VestergaardPeterson2006, Bentz2013, Wang2009, VestergaardOsmer2009, HoKim2015, Jun2015}, we focus our analysis on the comparison of our data with the only two other sample with the same characteristics, i.e. whose spectra were taken with the XShooter spectrograph and therefore cover a spectral range including all the broad lines of interest, Paper I and \cite{Mejia-Restrepo2016}. We notice that, while both these samples cover the same redshift range $z\sim1.5$, ours goes to higher redshift ($z\sim2.2$) and therefore can be interesting to make a comparison in terms of mass and Eddington ratio. Since the quasars in our sample are selected to be slightly brighter than those selected in Paper I, we expect them to be characterized by higher values for at least one of these two quantities.

\begin{figure*}[hbp!]
\centering
\includegraphics[scale=0.72]{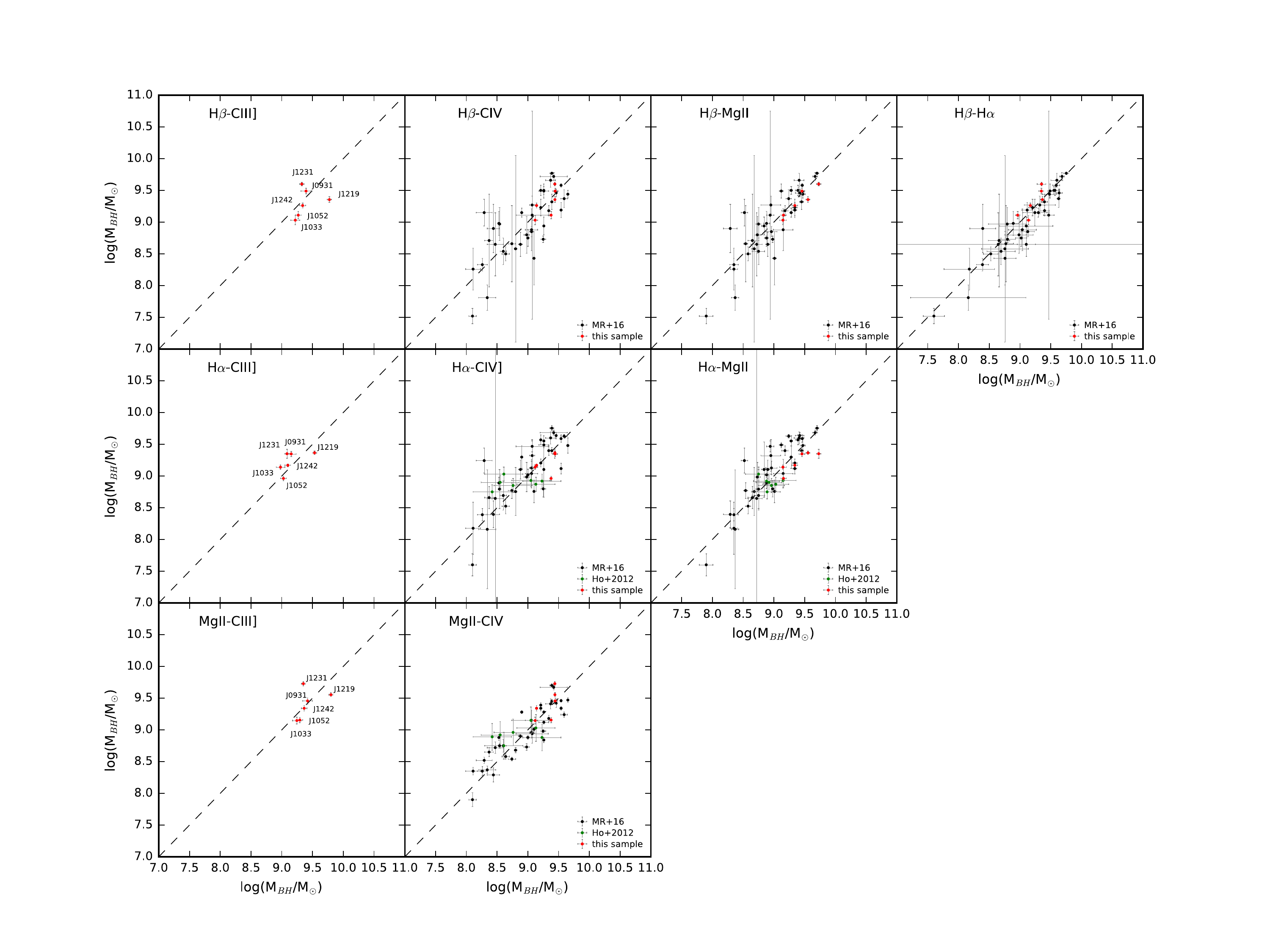}
\caption{ Correlations between virial masses. The lines labeled in the upper left of every panel refer to $Y$ and $X$ axis respectively. \\
$1^{st}$ column: correlations between \ciii]-based virial masses and masses derived from the other lines. The black dashed line represents the 1:1 relation. Intercept and scatter for all the relations are given in Tab.~\ref{tab6}.\\
$2^{nd}$, $3^{rd}$ and $4^{th}$ columns: 
M$_{BH}$ measurements for \citealt{Mejia-Restrepo2016} (black points), \citealt{Ho2012} (green points) and this work (red points) samples obtained with the prescriptions by \citealt{Mejia-Restrepo2016}.
Black dashed lines represent the 1:1 relation. The prescriptions we use are the \emph{local approach M$_{BH}$ calibrations, but corrected for the small systematic with respect to the global approach M$_{BH}$ calibrations} that \citealt{Mejia-Restrepo2016} present in the third column of their Tab.7. For Paper I only \civ, \mgii\ and H$\alpha$ are available. In the case of \civ, for Paper I sources, we had to use the $L_{1350}$ in place of the $L_{1450}$, because this is the closest specific luminosity available for this sample.
M$_{BH}$ estimates for our sources are given in Tab.~\ref{tab5}.}
\label{fig:Mbh_XShooter_all_samples}
\end{figure*}

In Fig. \ref{fig:Mbh_XShooter_all_samples} (second, third and fourth columns) we report the measurements of M$_{BH}$ obtained with the new prescriptions of \cite{Mejia-Restrepo2016} for our sample (see Tab.~\ref{tab5}) and for the Paper I sample, for which only the measurements pertaining to three lines out of four are present (in this work results for H$\beta$ are not included, given the poor signal to noise in this spectral range, and all the comparison are made with H$\alpha$). 
In the same figure we also show the \cite{Mejia-Restrepo2016} sample.
As for the H$\alpha$-based masses, we use their H$\alpha$ prescription with the luminosity at $5100$\AA, the same we used for the other samples. For all the lines we used the third column of \cite{Mejia-Restrepo2016} Tab.7, i.e. \emph{local approach M$_{BH}$ calibrations, but corrected for the small systematics with respect to the global approach M$_{BH}$ calibrations}. The \civ-based M$_{BH}$ for the \cite{Ho2012} sources are computed using $L_{1350}$ instead of $L_{1450}$, because this is the closest continuum luminosity available for this sample.
Our sample (red data points) fits very well in all cases and, on average, is located in the upper part of the global distribution.

We then compute the Eddington ratios for our sources and for \cite{Mejia-Restrepo2016} and Paper I samples, with the same prescription used in Paper 1 (bolometric luminosity from \cite{McLureDunlop2004} and Eddington luminosity $L_{Edd} = 1.26 \; 10^{38}(\textup{M}_{BH}/ \textup{M}_{\odot})\, \textup{erg s}^{-1}$) to verify if our sample is composed by higher accreting black holes. 
We evaluate the Eddington ratio for our objects both with H$\beta$ and with H$\alpha$ as virial estimators, while for \cite{Mejia-Restrepo2016} we recompute the H$\alpha, L_{5100}$ based values. In this way we can compare these values with those found  in Paper I, for which only H$\alpha$ measurements are available.

\begin{figure}
\centering
\includegraphics[scale=0.60]{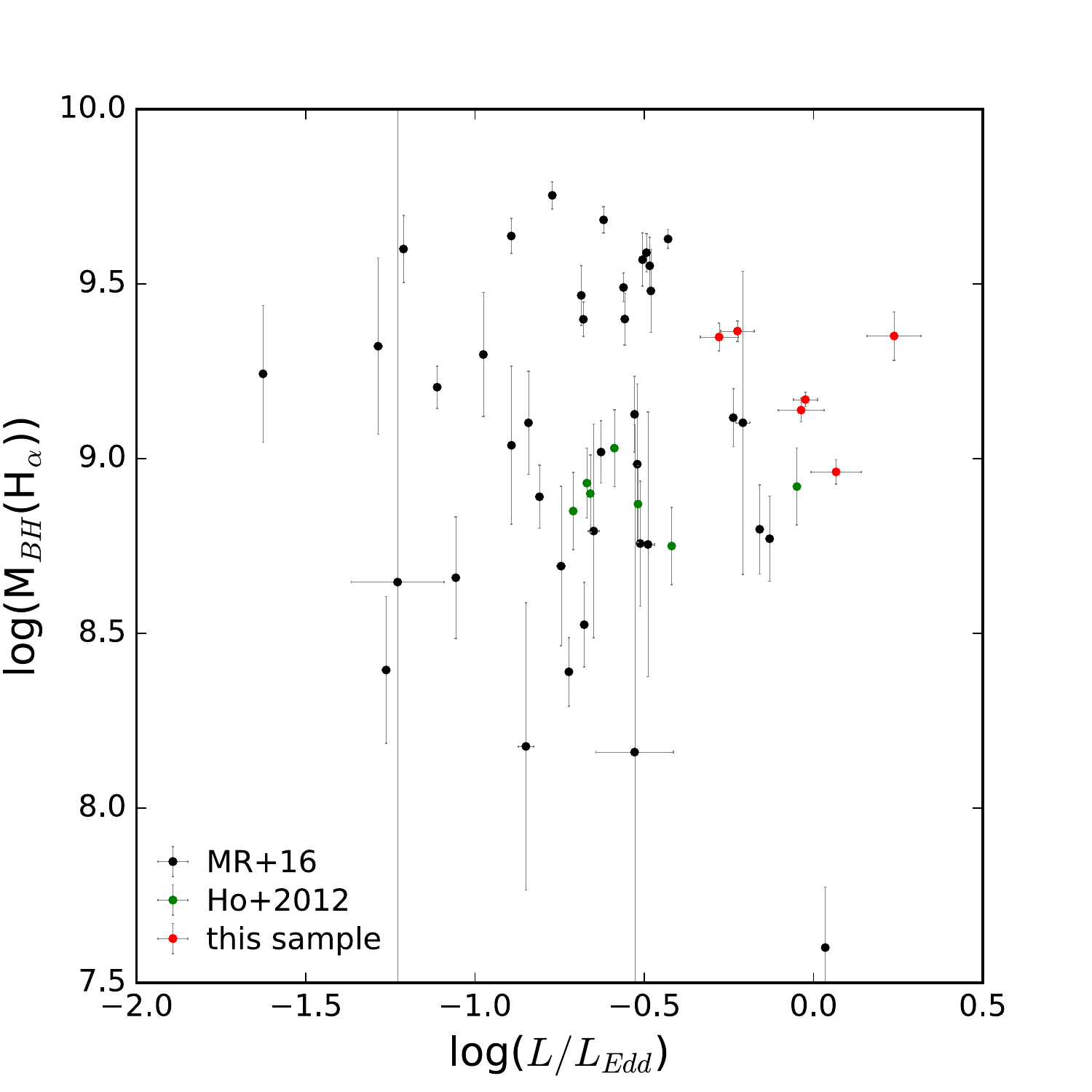}
\caption{M$_{BH}$ vs Eddington Ratios for \citealt{Mejia-Restrepo2016} (black points), \citealt{Ho2012} (green points) and this work (red points) sources. For the Eddington Ratios, we follow \citealt{McLureDunlop2004} for the bolometric luminosity and use $L_{Edd} = 1.26 \; 10^{38}(M_{BH}/M_{\odot}) erg\;s^{-1}$ for the Eddington luminosity.
We use the H$\alpha$-based M$_{BH}$, because this is the only Balmer line available for all the samples. The values for our objects are listed in Tab. \ref{tab5}.}
\label{fig:eddington_ratios}
\end{figure}

We find that our Eddington ratios (Tab. \ref{tab5}, Fig. \ref{fig:eddington_ratios}) are much higher on average than those of both Paper I and \cite{Mejia-Restrepo2016}.

The higher luminosity of our sample is therefore due both to the presence of more massive BHs and to the fact that they are accreting, on average, at higher rates.

\begin{table*}
\small
\caption{Mass estimations with \citealt{Mejia-Restrepo2016} prescriptions for our sample.}
\begin{center}
\begin{tabular}{l c c c c c c c c l }
    \hline
     
                                                 & J093147      & J103325 &     J105239    &   J121911      &   J123120    &     J124220 &      \\

\hline
log(M$_{\textup{\civ}}$/M$_{\odot}$)    &$9.44\pm0.04$&$9.12\pm0.04$&$9.38\pm0.02$&$9.44\pm0.01$& $9.44\pm0.01$ &$9.14\pm0.02$&\\
log(M$_{\textup{\mgii}}$/M$_{\odot}$)    &$9.46\pm0.05$&$9.15\pm0.06$&$9.16\pm0.04$&$9.55\pm0.03$& $9.73\pm0.03$ &$9.34\pm0.04$&\\
log(M$_{H\beta}$/M$_{\odot}$)  &$9.49\pm0.04$&$9.03\pm0.07$&$9.11\pm0.06$&$9.36\pm0.04$& $9.60\pm0.03$ &$9.26\pm0.04$&\\
log(M$_{H\alpha}$/M$_{\odot}$)&$9.35\pm0.04$&$9.14\pm0.03$&$8.96\pm0.04$&$9.37\pm0.03$& $9.35\pm0.07$ &$9.17\pm0.02$&\\ \relax
R$_{Edd}$ (H$\beta$ based)&$0.380\pm0.054$&$1.178\pm0.288$&$0.830\pm0.180$&$0.609\pm0.089$& $ 0.975\pm0.085$ &$0.763\pm0.097$&\\
R$_{Edd}$ (H$\alpha$ based)&$0.526\pm0.068$&$0.919\pm0.142$&$1.166\pm0.199$&$0.596\pm0.068$&$1.730\pm0.318$&$0.946\pm0.078$&\\
\hline
\end{tabular}
\label{tab5}
\end{center}
\end{table*}


\subsection{Can \textup{\ciii]} be used as a virial estimator?}
Unlike what found in Paper I, when looking at the \ciii] profile we do not recognize a different behaviour of the line with respect to the others (see Fig. \ref{all_lines_comparison}). We then decide to examine the relationships between the FWHM of \ciii] with those of the other lines.
Although \ciii] is not commonly used, some works analyse this line \citep{Greene2010, ShenLiu2012}. \cite{Greene2010} find only a slight correlation of \ciii] with \mgii\ FWHM, while \cite{ShenLiu2012} state that \ciii] linewidths correlates with \civ\ and therefore these lines could be emitted by the same region, then being characterized by the same issues (i.e. non virialization of the emitting region).
We find that this correlation (log(FWHM$_{\textup{\civ}}$)-log(FWHM$_{\textup{\ciii]}}$)) has a larger scatter with respect to those of H$\beta$ and H$\alpha$ and comparable with that of \mgii\ (see Tab. \ref{tab4} and Fig.\ref{fig:FWHM_comparison}, first column). However, the sample of \cite{ShenLiu2012} has much higher luminosity than ours and is composed by lower redshift sources. Moreover, they fit the line profiles on much narrower wavelenght ranges than ours. Additionally, for the line profile model they use two Gaussians tied to give a symmetric broad component for \ciii]. 
All these differences could then contribute to the discrepancy with their results.

In Fig. \ref{fig:FWHM_comparison} (first column) we notice the presence of only one outlier, J121911, for which the \ciii] complex appears to have a more ``boxy'' shape with respect to those the other sources, resulting in an more asymmetric \ciii] profile with an extended red wing (see the figures in Appendix for the results on the complete sample and Fig. \ref{all_lines_comparison} for a comparison of the line profiles). 
We have checked that including or excluding this point does not affect the fit and we decide to leave it in the sample.
The reason for this is that this point has larger errors, since the IDL routine we use to fit the linear relation (MPFITEXY, \citealt{Williams2010}, based on the MPFIT package \citep{Markwardt2009}) considers the errors in both $x$ and $y$ variables.
J121911 does not stand out evidently in the case of the line dispersion (Fig. \ref{fig:sigma_comparison}, first column), but for the same reason we do not consider it as a reliable measurement.
Results of fits assuming a linear correlation between the quantities are listed in Tab. \ref{tab4}.

\ciii] is not usually mentioned among the possible virial estimators. This is mostly due to (1) its smaller intensity and (2) to the blend with other emission lines in the same complex.
The use of only one component to fit the broad lines, instead of two or more Gaussians, is more robust against the degeneracy in the profile fitting, also thanks to the good quality of our data.
The \civ\ line should not be used in virial estimates because, although it is more intense, it is contaminated by non-virial components.
On the contrary, the preliminary line comparison (Fig. \ref{all_lines_comparison}) shows that \ciii] behaves very similarly to the lines that are mostly virialized (H$\beta$, H$\alpha$ and \mgii). Therefore, we attempt to find a virial relationship for \ciii] comparing measurements for this line with virial masses based on the other lines.

Given the smallness of our sample, in addition to a fixed dependence of the M$_{BH}$ on the velocity of the emitting gas according to the virial assumption, we also fix the dependence on the luminosity as $\textup{M}_{BH} \propto L^{0.5}$. This choice is perfectly consistent with the hypothesis of photoionization in the BLR and with what found in previous works \citep{Bentz2013}.
The BH mass is then given by the equation
\begin{equation} \label{eq:Mbh_CIII}
\textup{M}_{BH} = C\; \textup{FWHM}^{2} \, L^{0.5}\;,
\end{equation}
where the only free parameter is the scaling factor $C$.
We chose to use the luminosity at $1450$\AA, as the closest to the \ciii] line that we can measure in a continuum window reasonably free by other emissions.
In Tab. \ref{tab6} we report our results (scaling factor $C$ and scatter $\Delta$) for the comparison of the \ciii] based M$_{BH}$ with those from all the other lines except \civ\ (Fig. \ref{fig:Mbh_XShooter_all_samples}, first column). Since \civ, if not corrected, does not share the property of virialization of the emitting region, we do not consider this line for this comparison. Tab.~\ref{tab6} also shows the M$_{BH}$ estimates for all the sources derived from the \ciii] line. The scatter in these relations is comparable (only in the case of M$_{BH}$(\ciii])-M$_{BH}$(H$\beta$) is larger) with those of the mass relations involving \civ\ ($0.14$, $0.15$ and $0.16$ dex with the only six objects in our sample and $0.23$, $0.21$ and $0.20$ dex for the whole sample, for H$\beta$, H$\alpha$ and \mgii, respectively). However, the strong similarity of the \ciii] profile with the lines emitted by virialized gas (H$\beta$ and \mgii) suggests that, at least for this sample, we can use this line as a virial estimator. 
The use of the double power law function as a model to fit broad components helps in removing the degeneracy in the \aliii, \siiii] and \ciii] complex and, therefore, in retrieving the \ciii] profile more accurately. Despite the scatter in the \ciii] virial relationships is as large as that of \civ, this line does not seem to be affected by contamination by not-reverberating components.
Of course this sample is composed only of six sources, one of which (J121911) represents an outlier for what concerns the \ciii] behaviour, therefore a significantly enlarged sample is needed to ensure the goodness of this line as a virial estimator. Moreover, due to the strong blending of the $1900$\AA~ complex, spectra of very high quality and S/N are required in order to disentangle the different emission components. This fact can therefore limit the use of this line but, in principle, \ciii] seems to represent a better choice with respect to \civ\ as it is.
Since we are comparing our \ciii] based masses with those obtained using the \cite{Mejia-Restrepo2016} prescription and since they do not provide virial relationships for $\sigma$, we limit our analysis to the FWHM.

\begin{table*}[htbp!]
\small
\caption{Correlations of \ciii] based virial masses with the other lines. 
We fit our data points to the relation $\textup{log}(\textup{M}_{BH_{line}})=C + 2 \textup{log}(\textup{FWHM}_{\textup{\ciii]}}) + 0.5 \log(L_{1450\AA})$ for every line pair and find the only free parameter $C$ and the scatter for the relations, given in the first two columns of the table. We also report virial estimates for all the sources with these prescriptions.}
\begin{center}
\renewcommand{\arraystretch}{1.2}
\begin{tabular}{l c c c c c c c c c  l }
    \hline
    & \multicolumn{2}{c}{whole relation} & \multicolumn{6}{c}{log(M$_{BH}$/M$_{\odot}$) individual sources} \\

                       &$C$ (km/s)               &$\Delta$ (dex) &  J093147  & J103325 & J105239 & J121911 & J123120 & J124220   \\
\hline
\ciii]-H$\beta$ &$-6.87\pm0.02$ &$0.27$&$ 9.39\pm0.08$ &$9.22\pm0.07$ &$9.27\pm0.05$ &$9.77\pm0.03$ &$9.33\pm0.03$ & $9.34\pm0.05$\\
\ciii]-H$\alpha$& $-6.63\pm0.02$& $0.14$&$9.15\pm0.08$ &$8.98\pm0.07$ &$9.03\pm0.04$ &$9.53\pm0.03$ &$9.08\pm0.03$ &$9.10\pm0.04$\\
\ciii]-\mgii\          & $-6.90\pm0.02$& $0.20$&$9.42\pm0.08$ &$9.24\pm0.07$ &$9.29\pm0.05$ &$9.80\pm0.03$ &$9.35\pm0.03$ &$9.36\pm0.05$    \\
\hline
\end{tabular}
\label{tab6}
\end{center}
\end{table*} 


\section{Summary}

We examined a sample of six quasars at redshift $z{\sim}2.2$, whose spectra were taken with the XShooter spectrograph. This instrument covers a very large spectral range, allowing the simultaneous comparison of the most used virial estimators.
We compare our results with those of the only two other samples observed with XShooter (Paper I and \citealt{Mejia-Restrepo2016}).
The analysis gives the following results:
\begin{enumerate}
\item[1.] The comparison of the line profiles shows that H$\beta$, H$\alpha$ and \mgii\ behave in a similar way, as expected for virialized gas.
\item[2.] \civ\ is by far the line that deviates most from this condition because of its strong blueshifts and asymmetry.
\item[3.] We find \ciii] to behave consistently with the other lines, in contrast to \civ.
\item[4.] Comparisons of the linewidths obtained for every line give a similar scatter for the FWHM and the line dispersion $\sigma$. However, we chose to focus our analysis on FWHM, to be consistent with the works we are comparing our sample to.
\item[5.] We compute virial masses for our sample and for the sources in Paper I, using the prescriptions by \cite{Mejia-Restrepo2016}. All the sources follow the relations. A comparison with Paper I and \cite{Mejia-Restrepo2016} shows that our higher redshift sample has larger M$_{BH}$ and higher accreting rates.
\item[6.] Notwithstanding the smallness of our sample, we suggest a new virial mass prescription based on the FWHM of \ciii], which can be considered as a valid substitute of \civ\ for sources in which only this spectral window is present, if a high quality spectrum is available and a proper modelization of the \feii\ and \feiii\ emissions is included in the analysis.
Unlike \civ\ in fact, this line seems to share the behaviour of the lines emitted by virialized gas.
\end{enumerate}

\FloatBarrier

\vspace{0.5cm}

\begin{acknowledgements}
The authors would like to thank the anonymous referee for helpful comments and suggestions that considerably improved the work.
They also thank Alvaro Alvarez for support during the observations and Marianne Vestergaard for kindly providing the I Zw 1 iron templates from \cite{VestergaardWilkes2001}.
LCH acknowledges support from the Chinese Academy of Science (grant No. XDB09030102), National Natural Science Foundation of China (grant No. 11473002), and Ministry of Science and Technology of China (grant No. 2016YFA0400702).
GP acknowledges the Bundesministerium f\"{u}r Wirtschaft und 
Technologie/Deutsches Zentrum f\"{u}r Luft- und Raumfahrt 
(BMWI/DLR, FKZ 50 OR 1408 and FKZ FKZ 50 OR 1604) 
and the Max Planck Society.

 \end{acknowledgements}


\bibliographystyle{aa}
\bibliography{bib_XS_pap}

\begin{appendix}
\section{Complete sample fits}

\begin{figure*}
   {\includegraphics[scale=0.165]{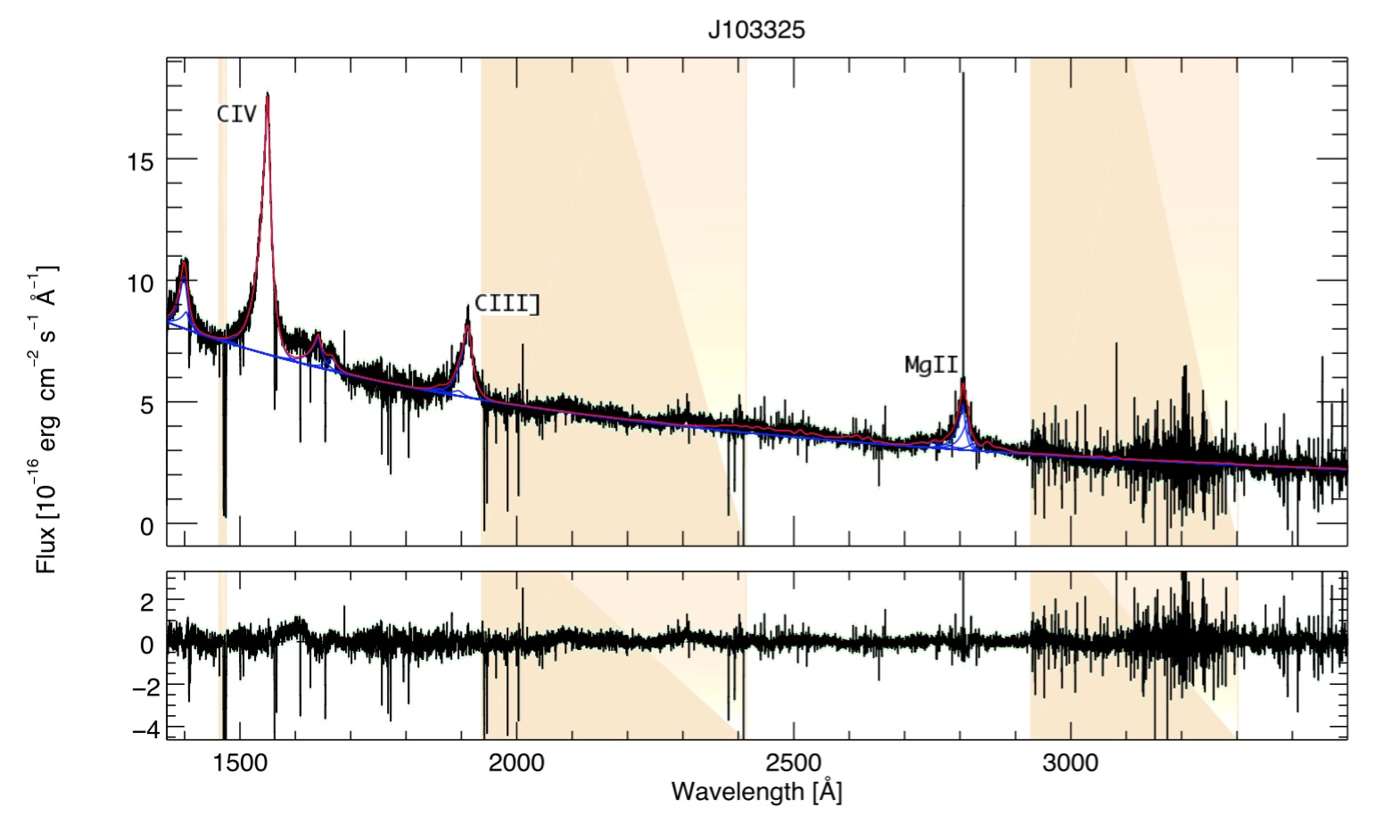}}
   {\includegraphics[scale=0.165]{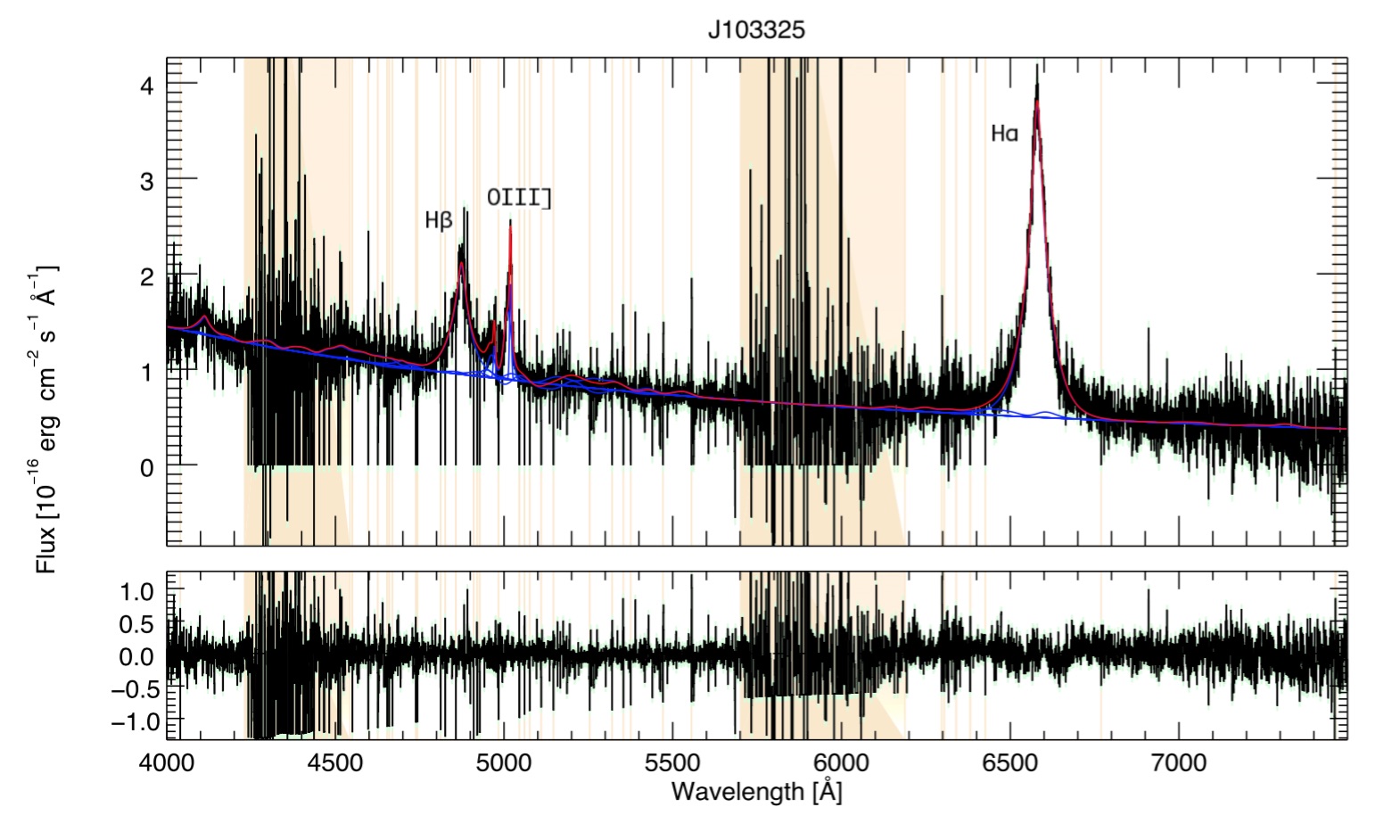}}
   {\includegraphics[scale=0.165]{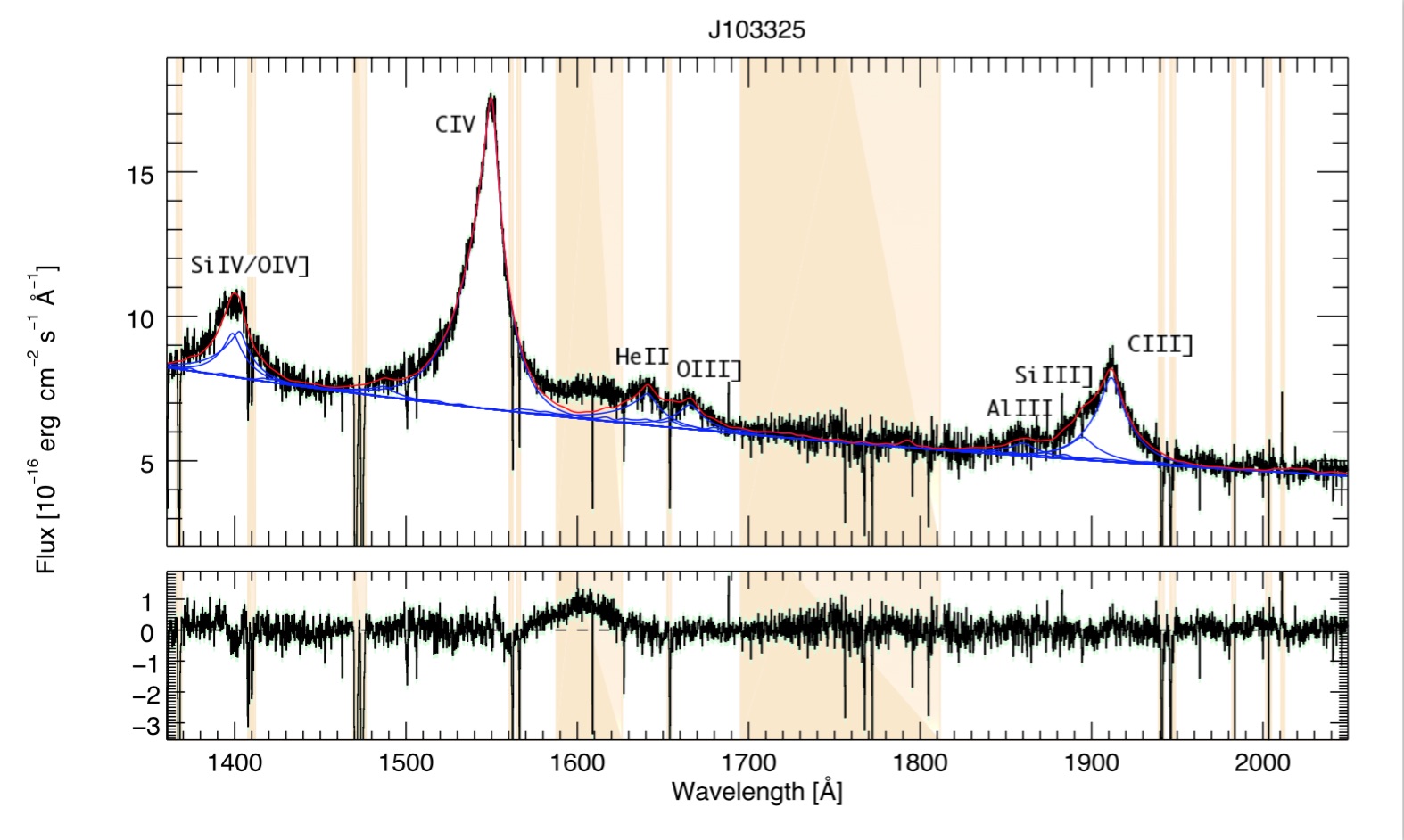}}
   {\includegraphics[scale=0.165]{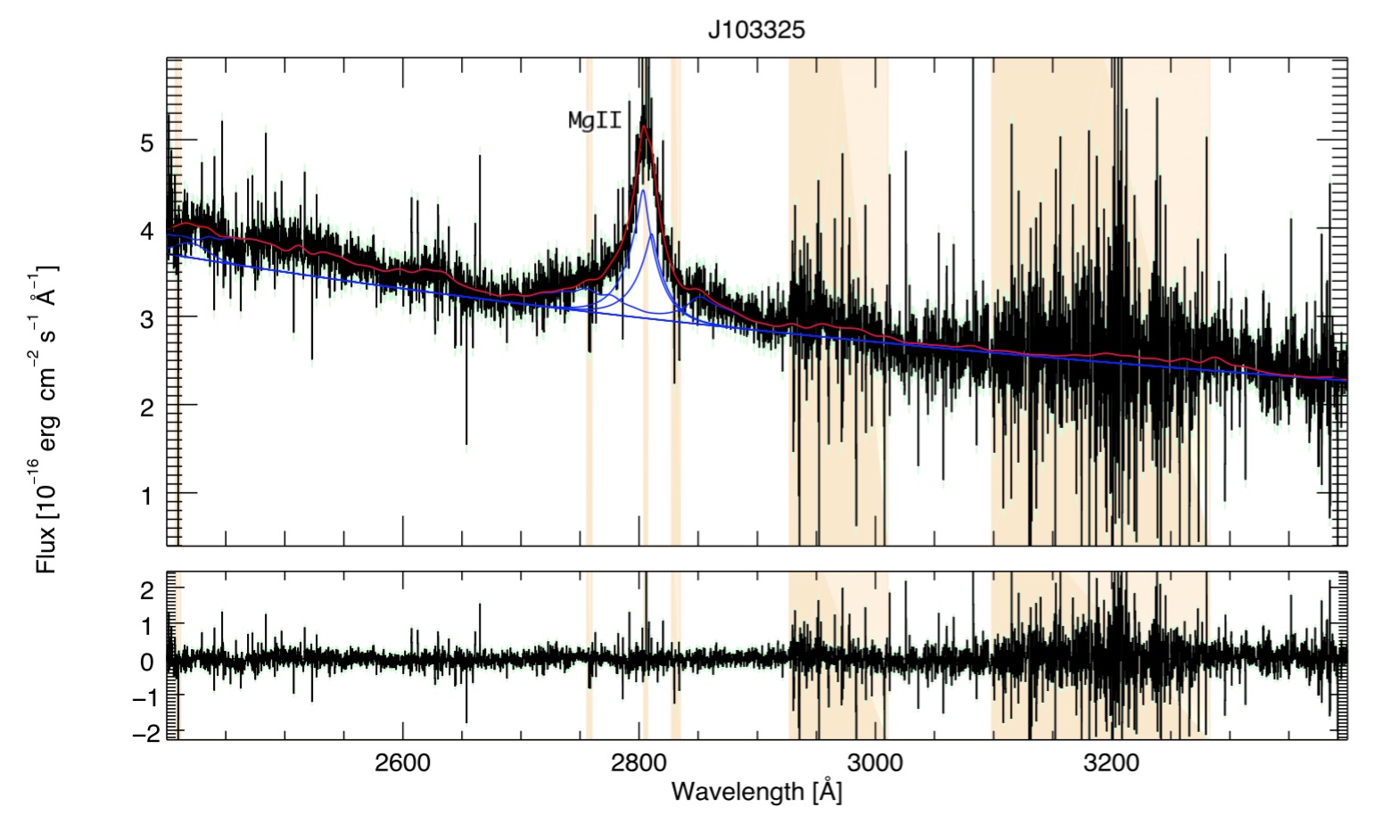}}
   {\includegraphics[scale=0.26]{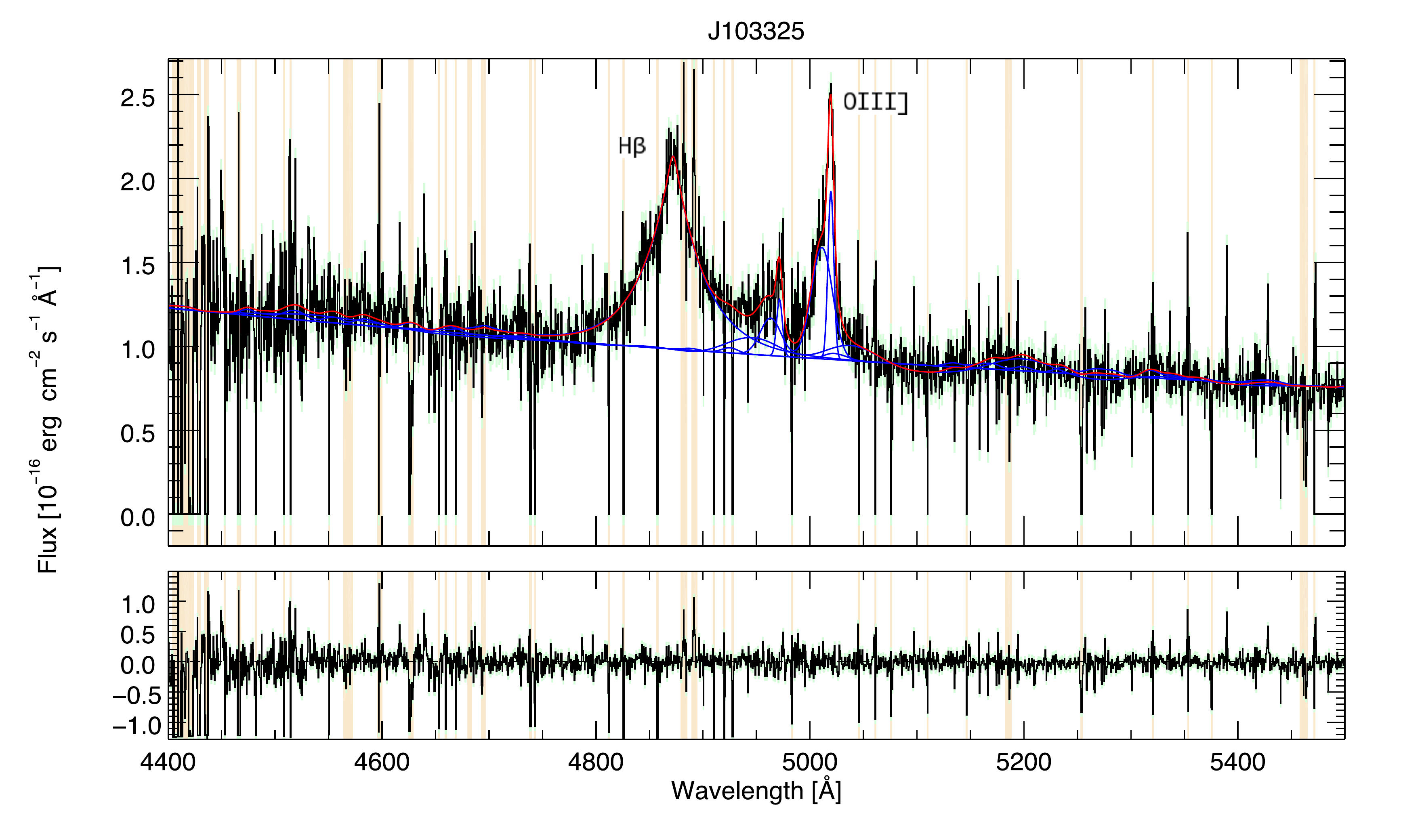}}
   {\includegraphics[scale=0.26]{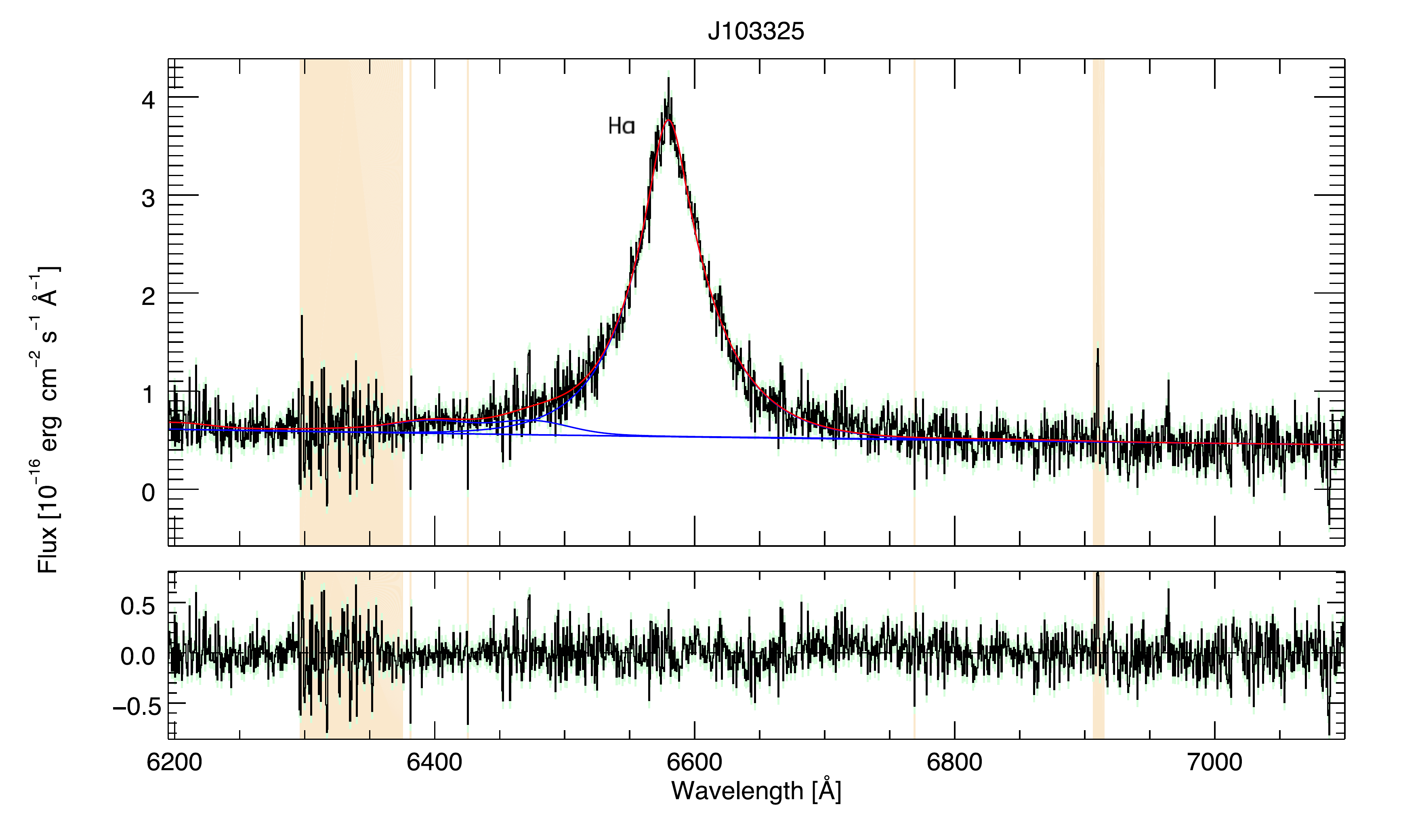}}
 \caption{Fits for all the examined spectral windows (\civ-\ciii]-\mgii\ and H$\beta$-H$\alpha$ large windows, \civ-\ciii], \mgii, H$\beta$ and H$\alpha$ small windows) for J103325. The black line is the original spectrum. The blue solid lines are the best fit models for the emissions (continuum, \feii\ and emission lines) and the red solid line is the total best fit. The lower panels show the residuals between the best fit model and the original spectrum.
The colored regions are those we chose to mask. This choice can be due to the presence of strong emission blending, to a non representativeness of the \feii\ templates or, in general, to a lack of knowlegde about what kind of emission is able to reproduce such features and the presence of noise.}
 \label{fig:fits_figures_J103325}
\end{figure*}
\begin{figure*}
   {\includegraphics[scale=0.165]{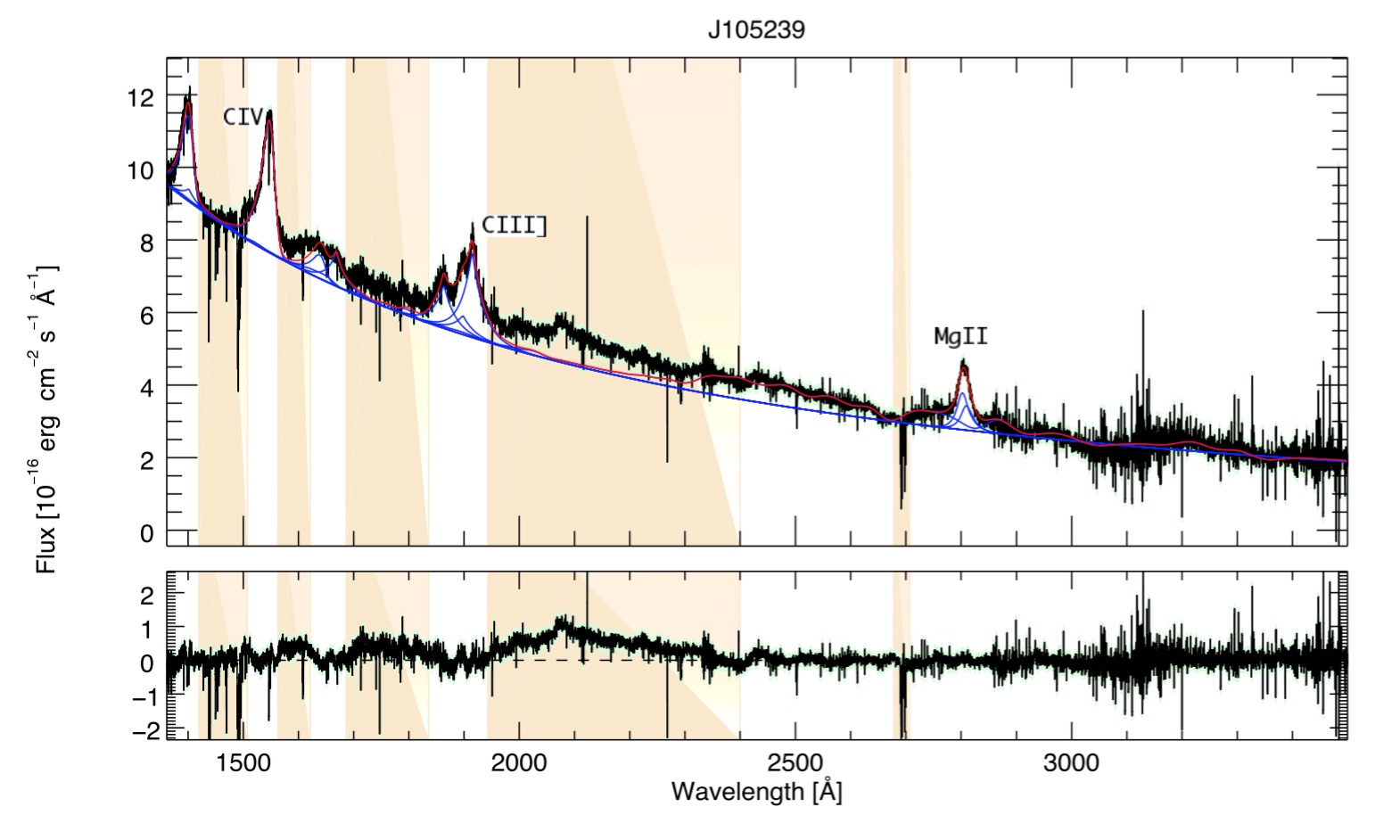}}
   {\includegraphics[scale=0.165]{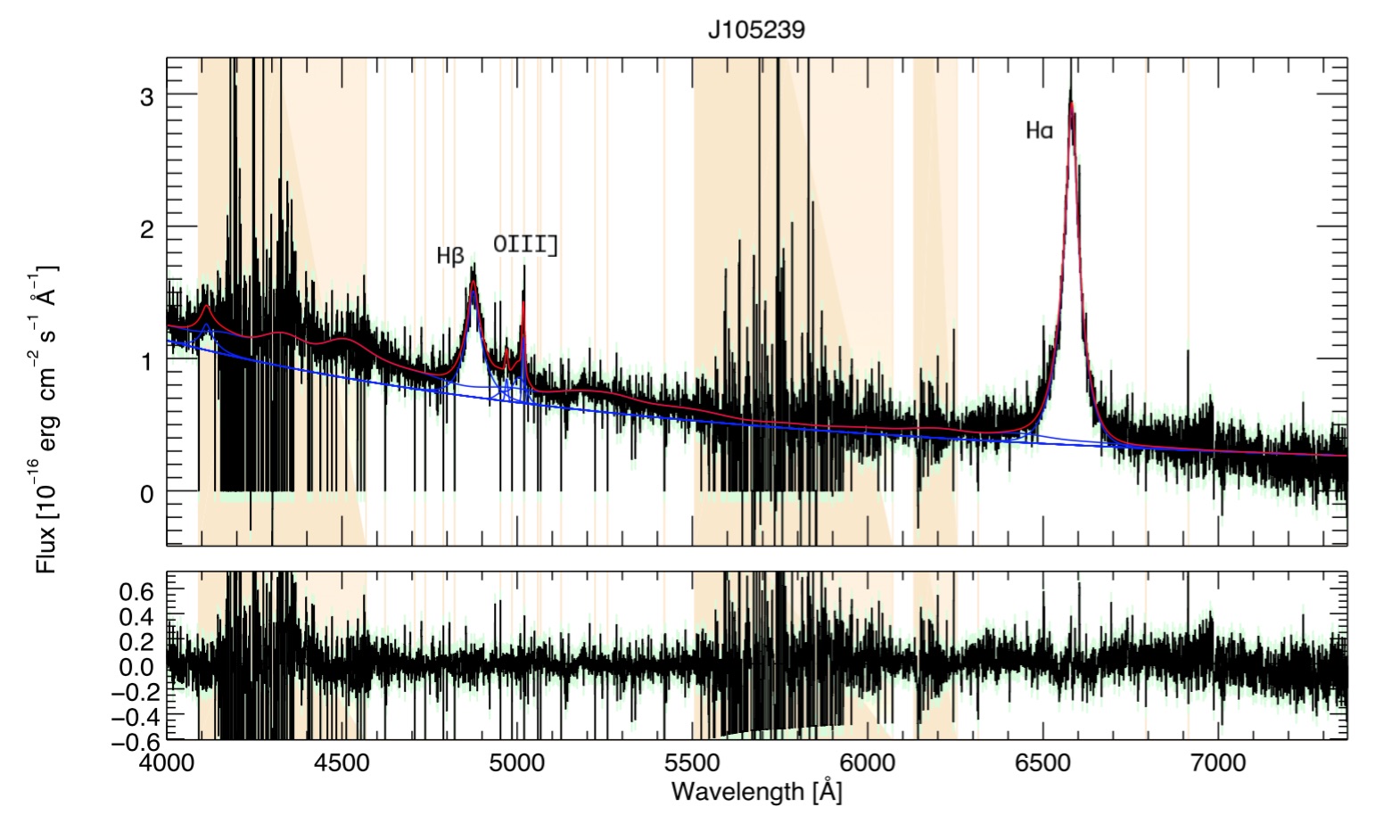}}
   {\includegraphics[scale=0.165]{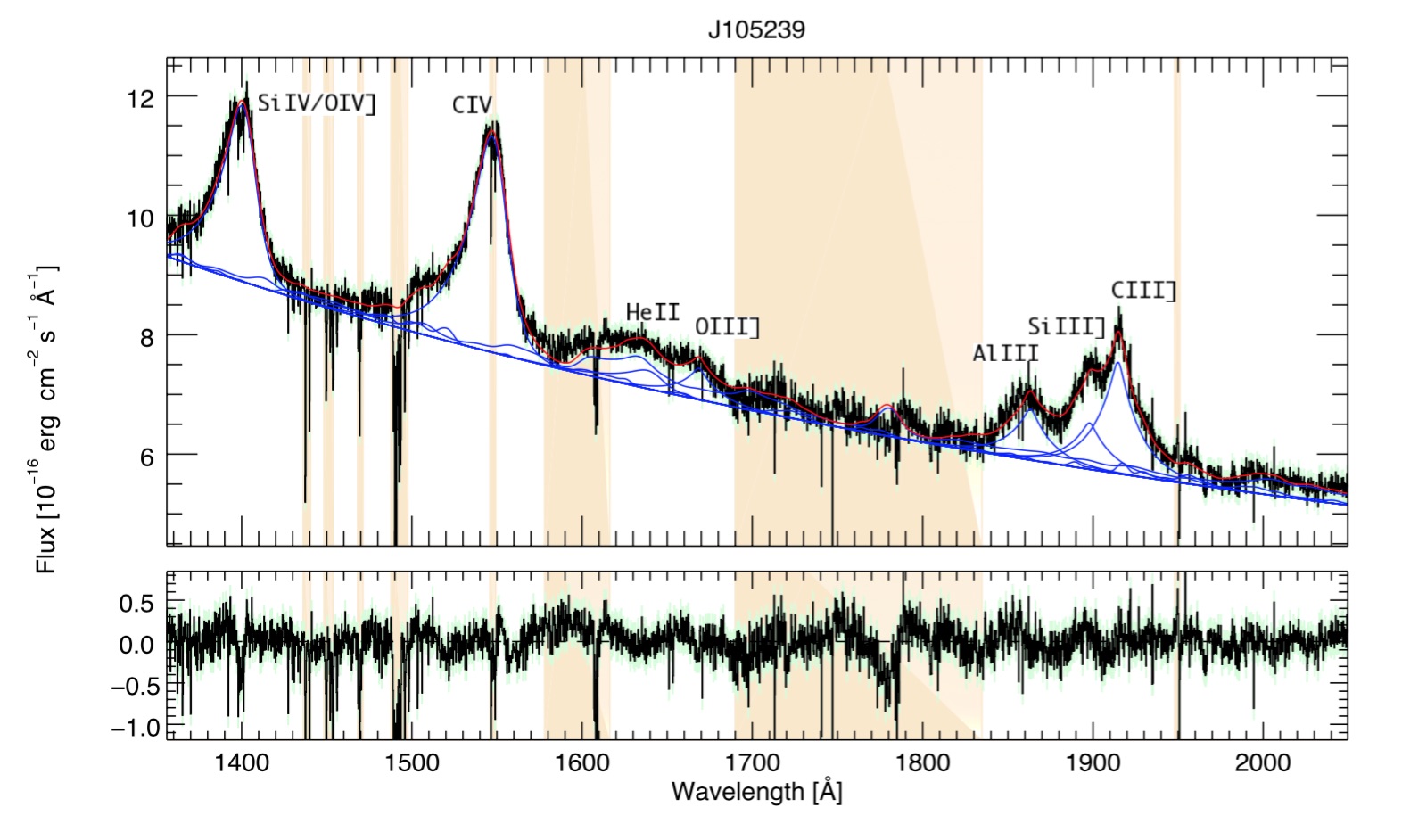}}
   {\includegraphics[scale=0.165]{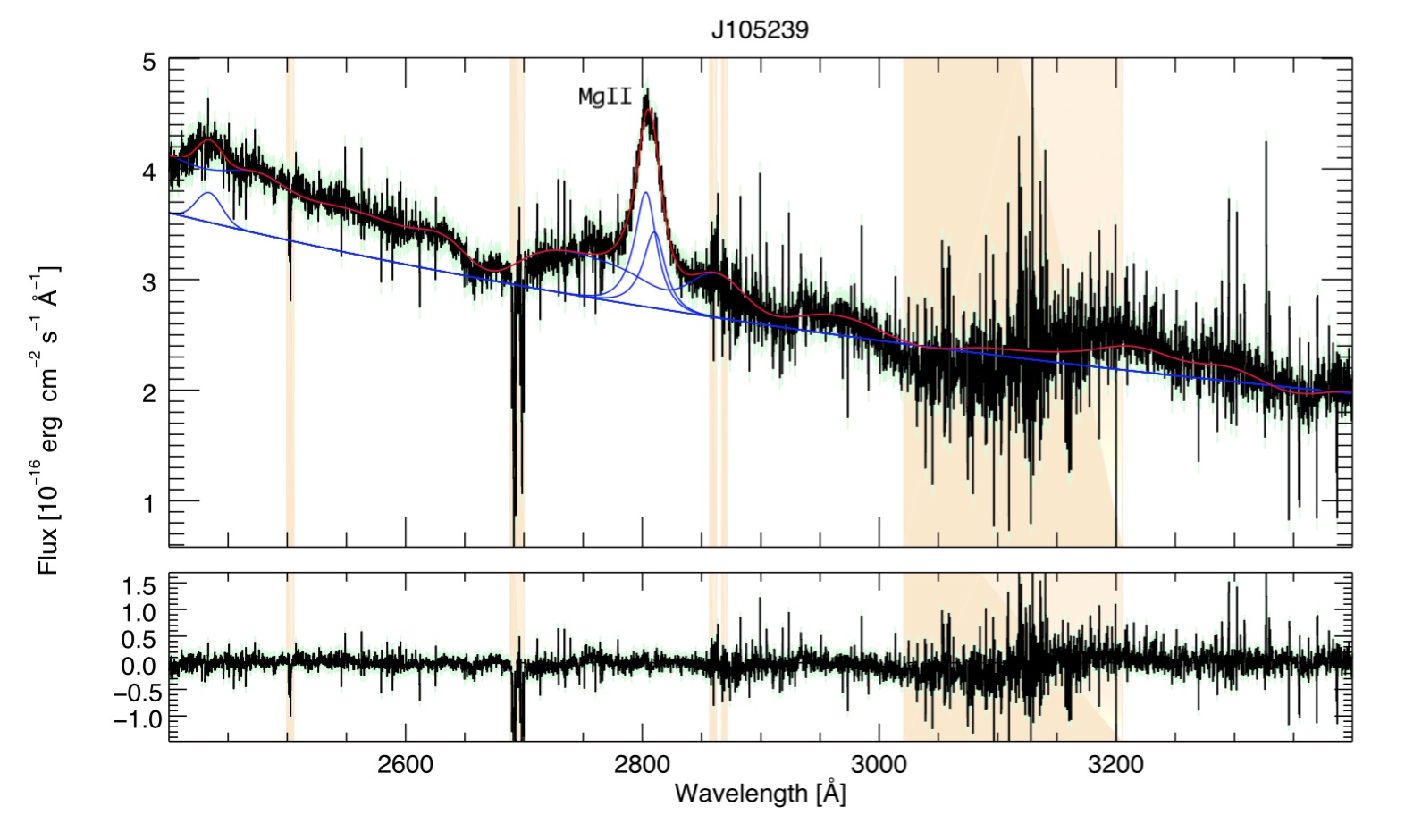}}
   {\includegraphics[scale=0.26]{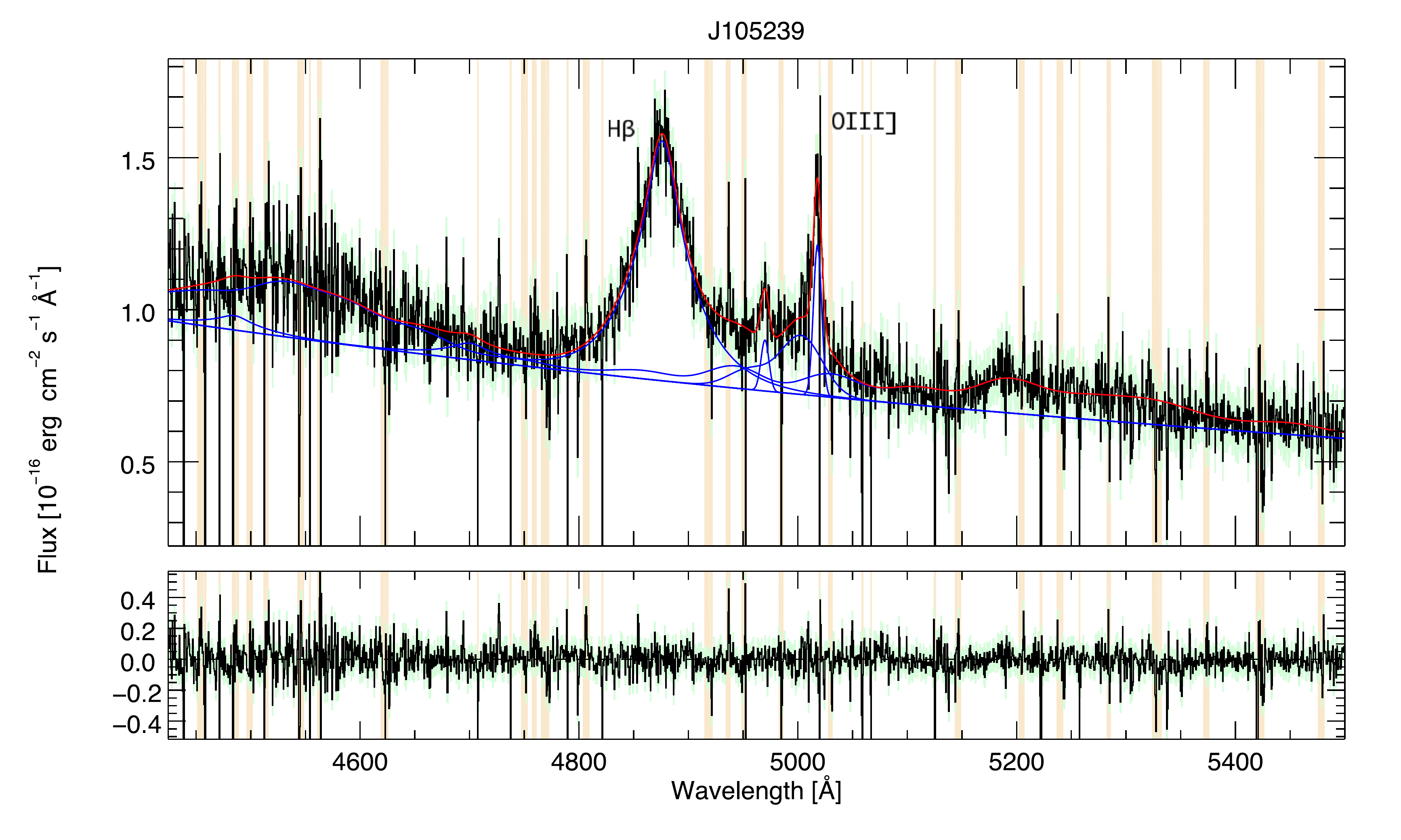}}
   {\includegraphics[scale=0.26]{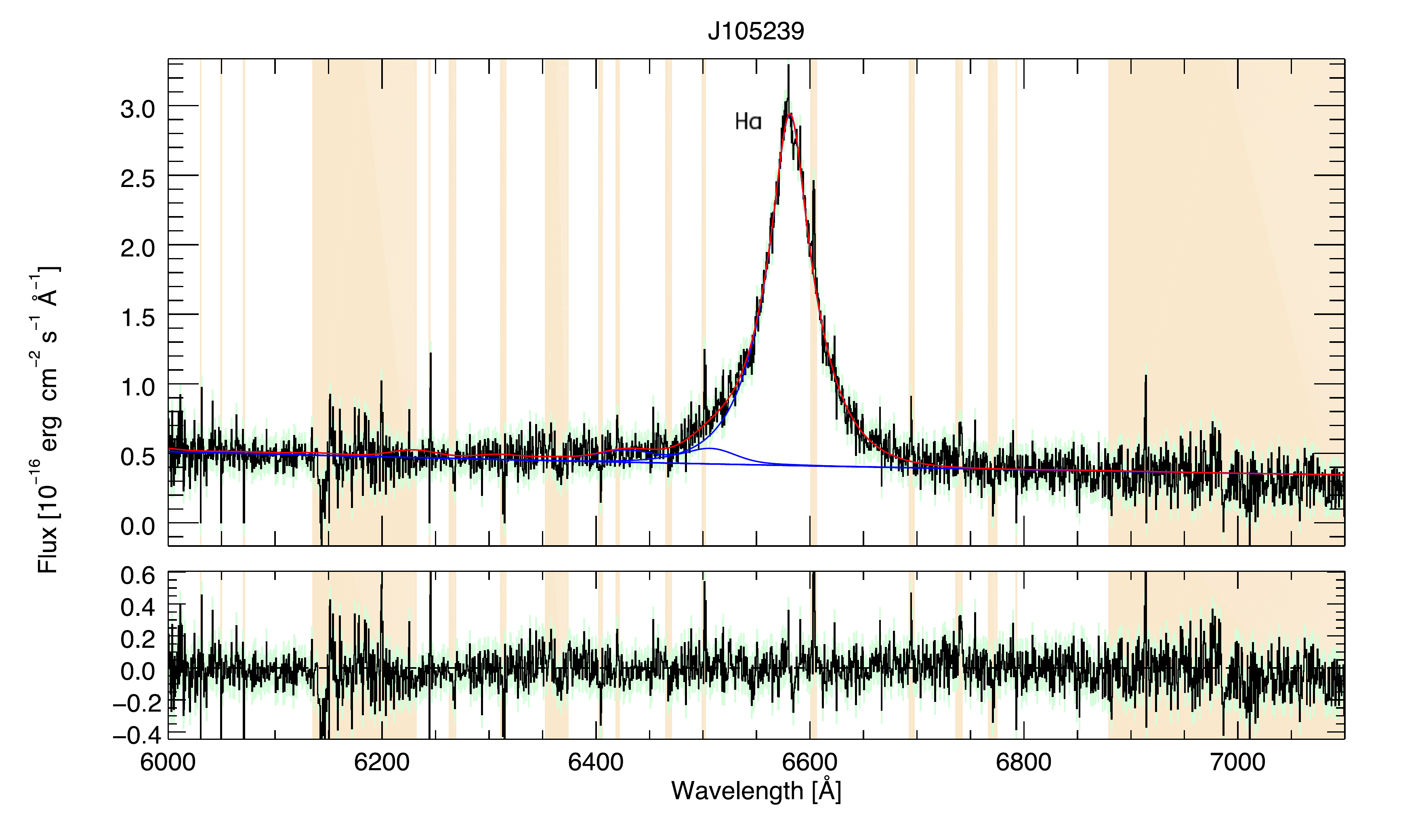}}
 \caption{Same as in the figure above but for J105239.}
 \label{fig:fits_figures_J105239}
\end{figure*}
\begin{figure*}
   {\includegraphics[scale=0.165]{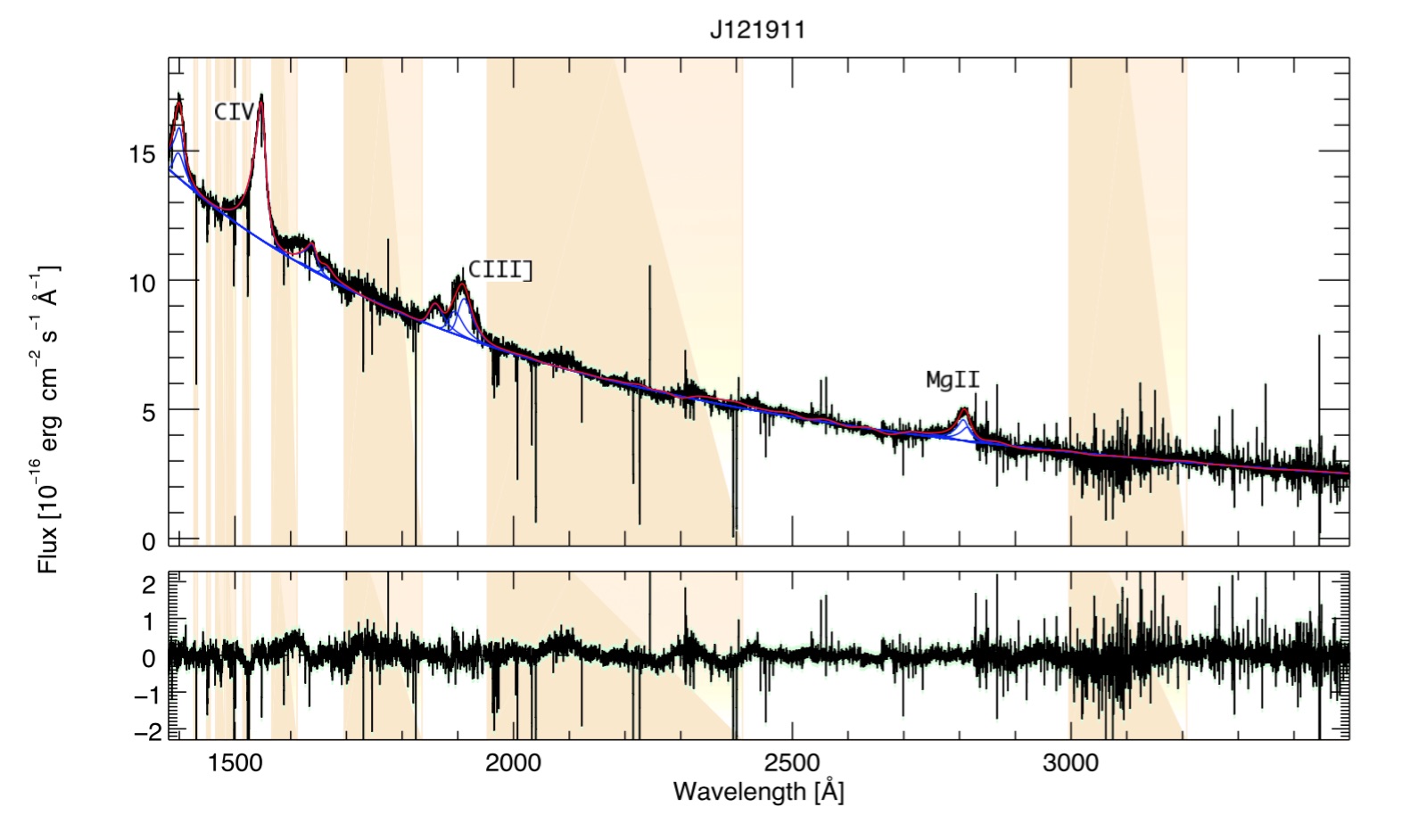}}
   {\includegraphics[scale=0.165]{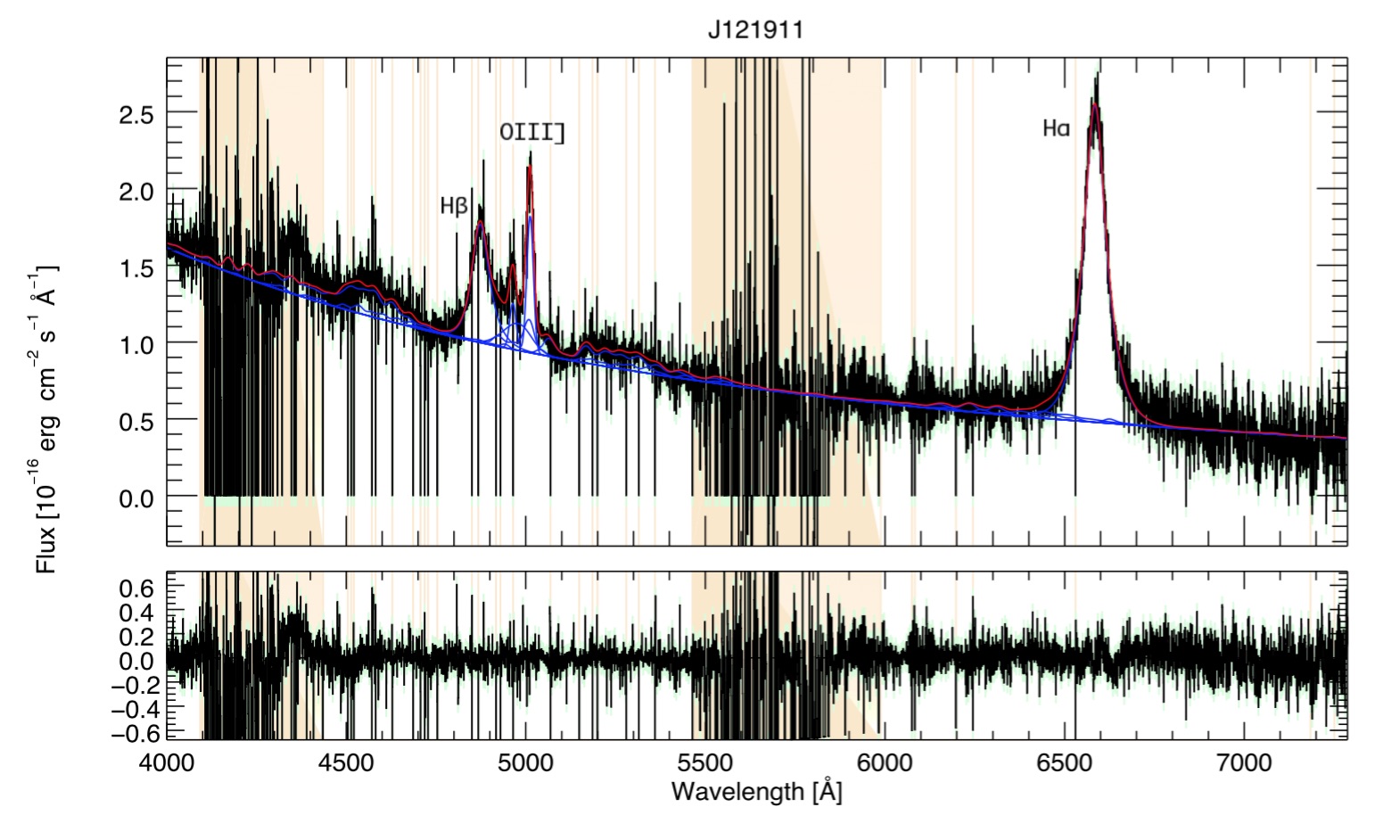}}
   {\includegraphics[scale=0.165]{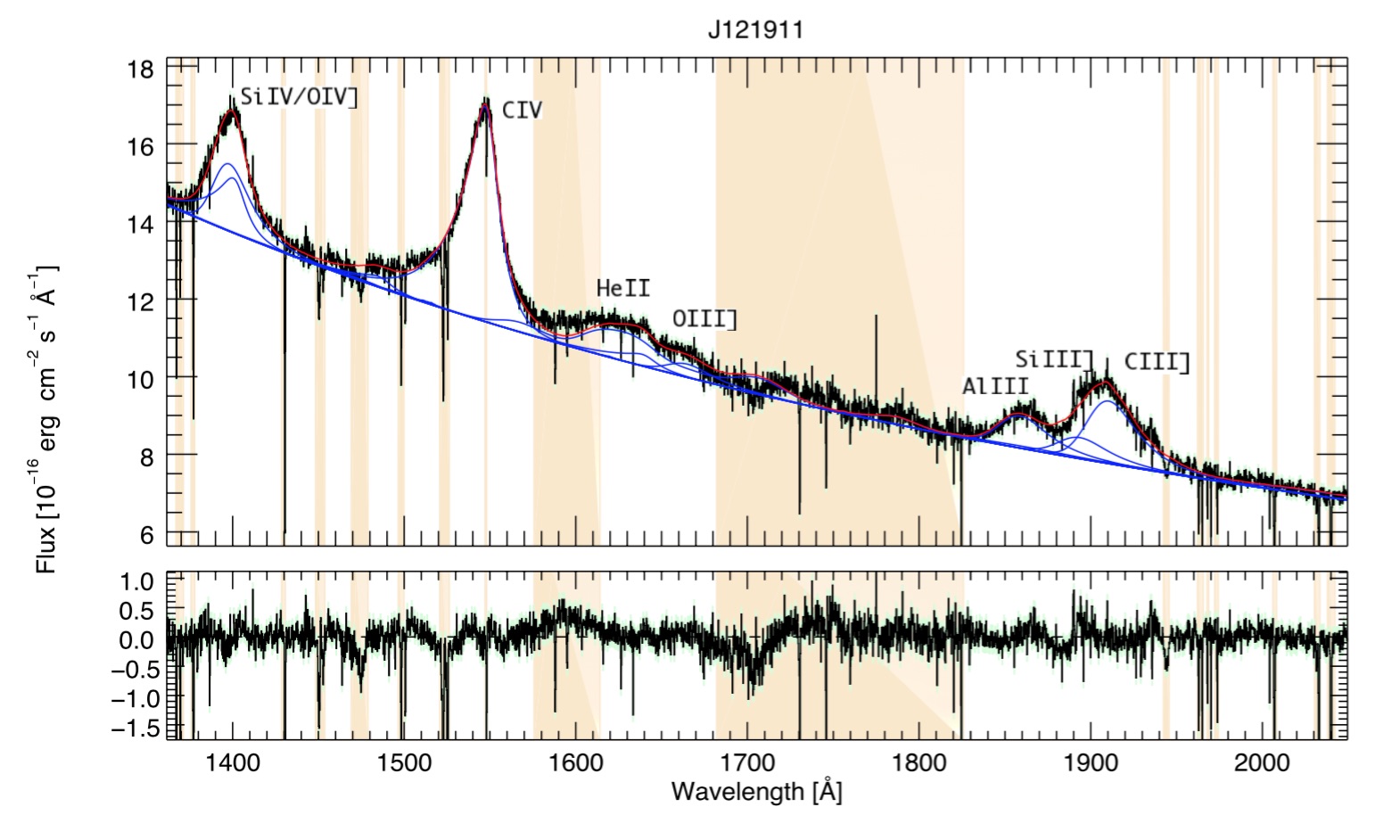}}
   {\includegraphics[scale=0.165]{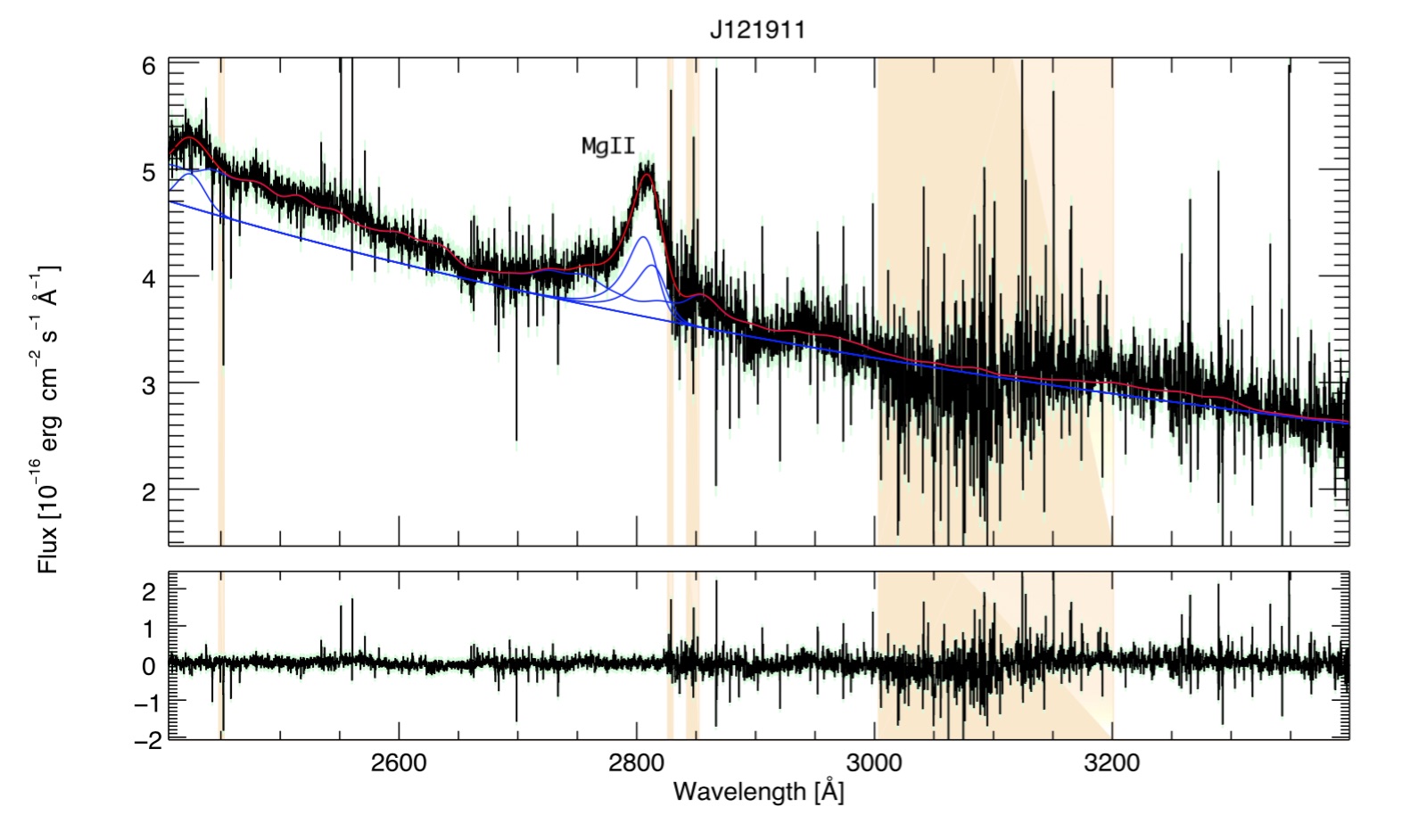}}
   {\includegraphics[scale=0.26]{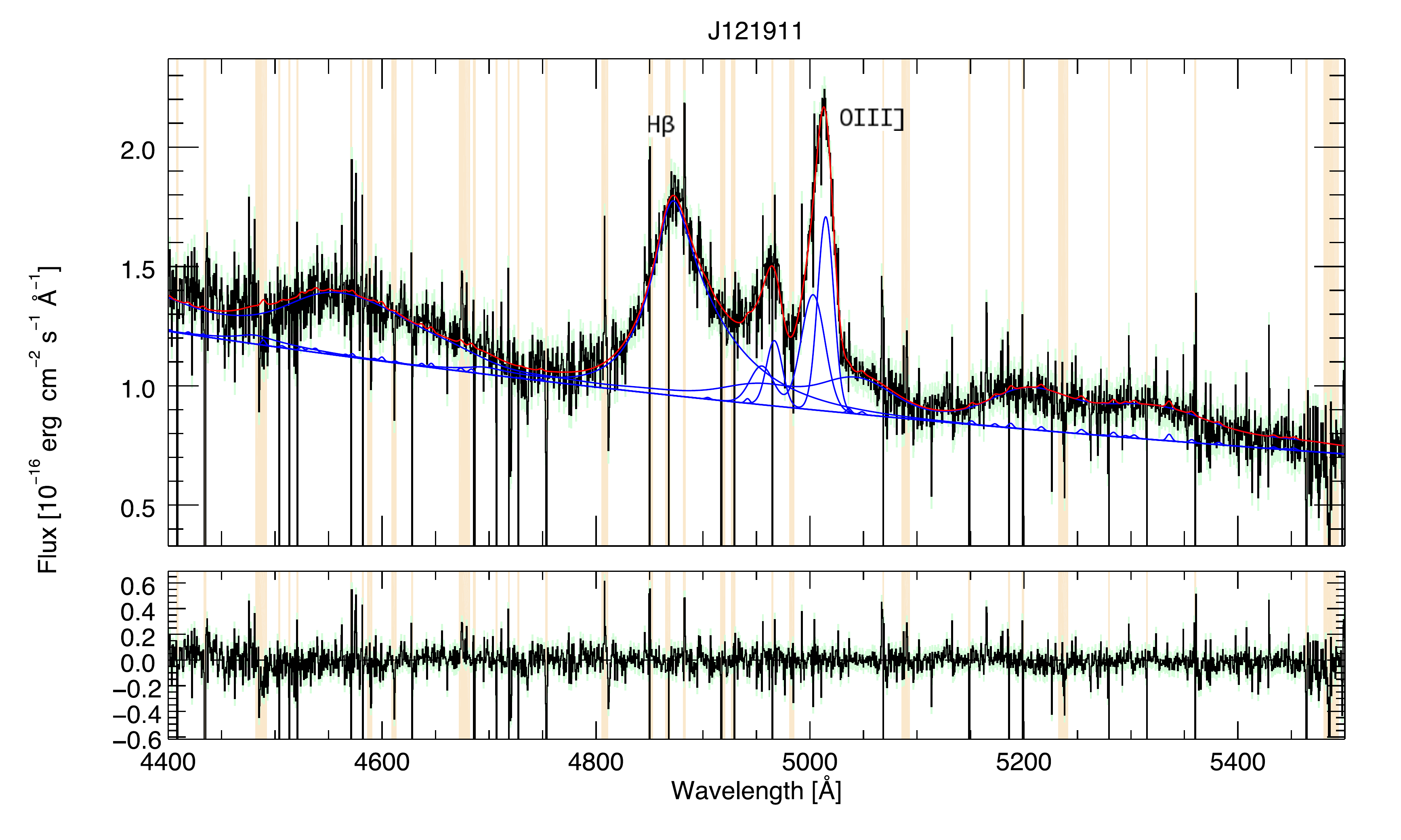}}
   {\includegraphics[scale=0.26]{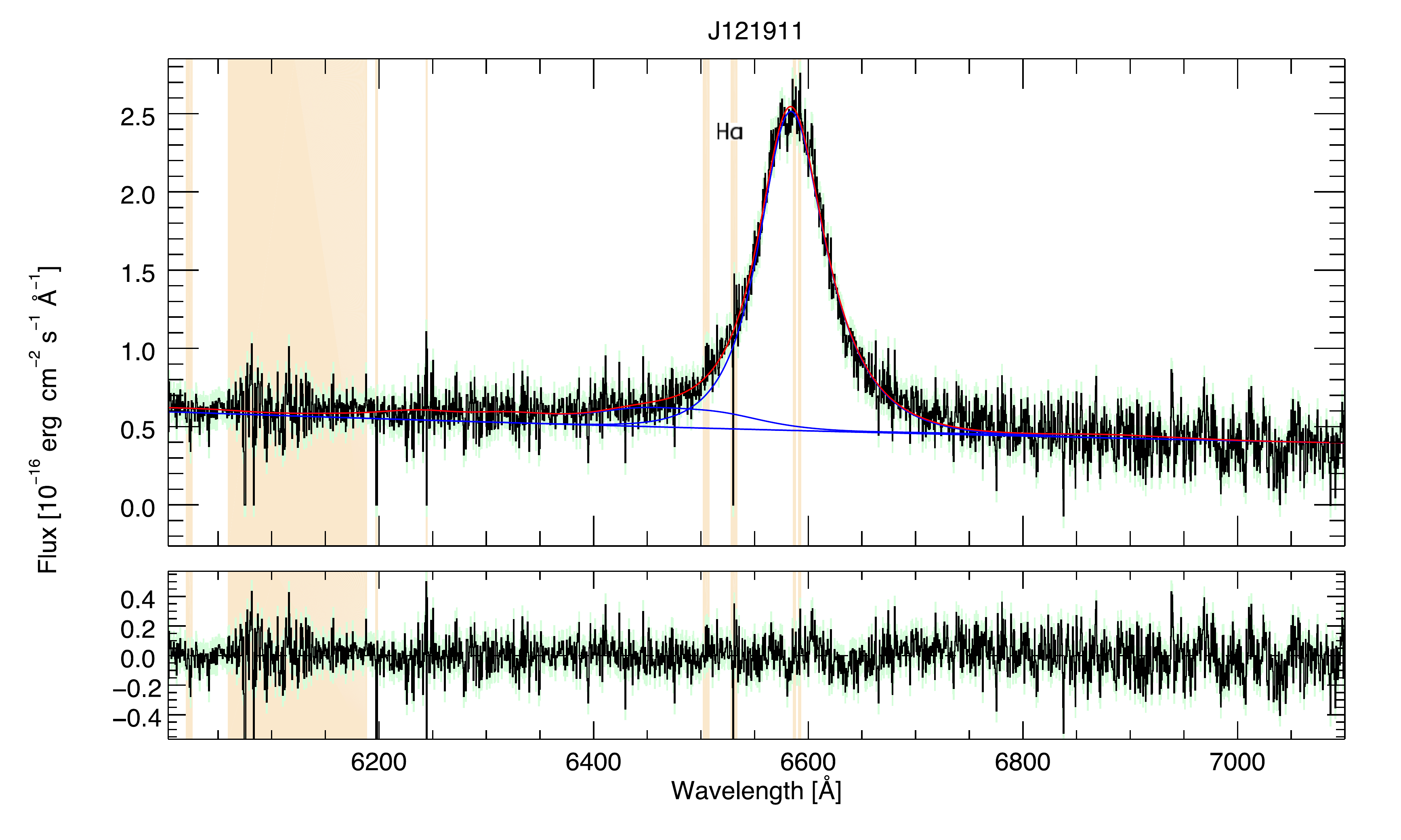}}
 \caption{Same as in the figure above but for J121911.}
 \label{fig:fits_figures_J121911}
\end{figure*}
\begin{figure*}
   {\includegraphics[scale=0.165]{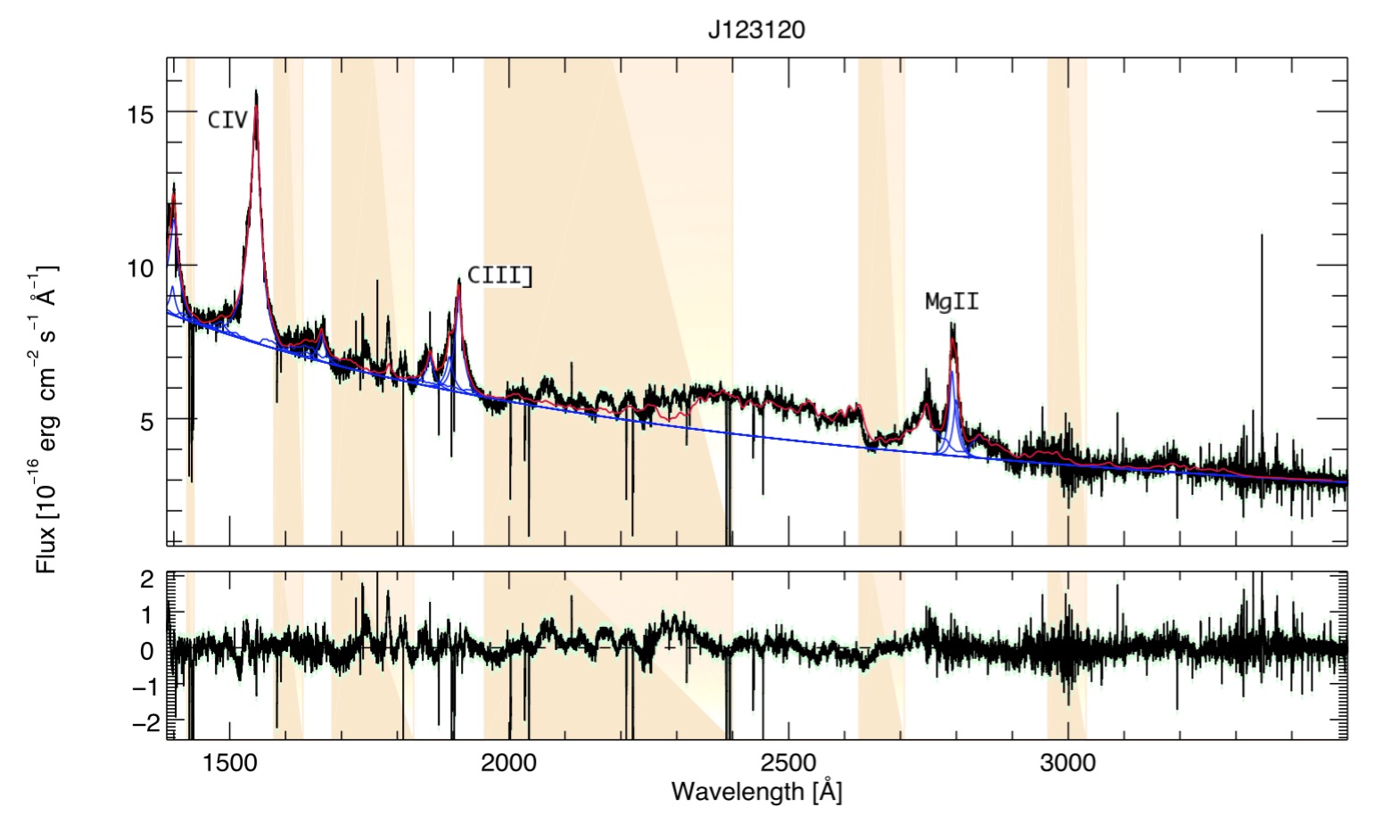}}
   {\includegraphics[scale=0.165]{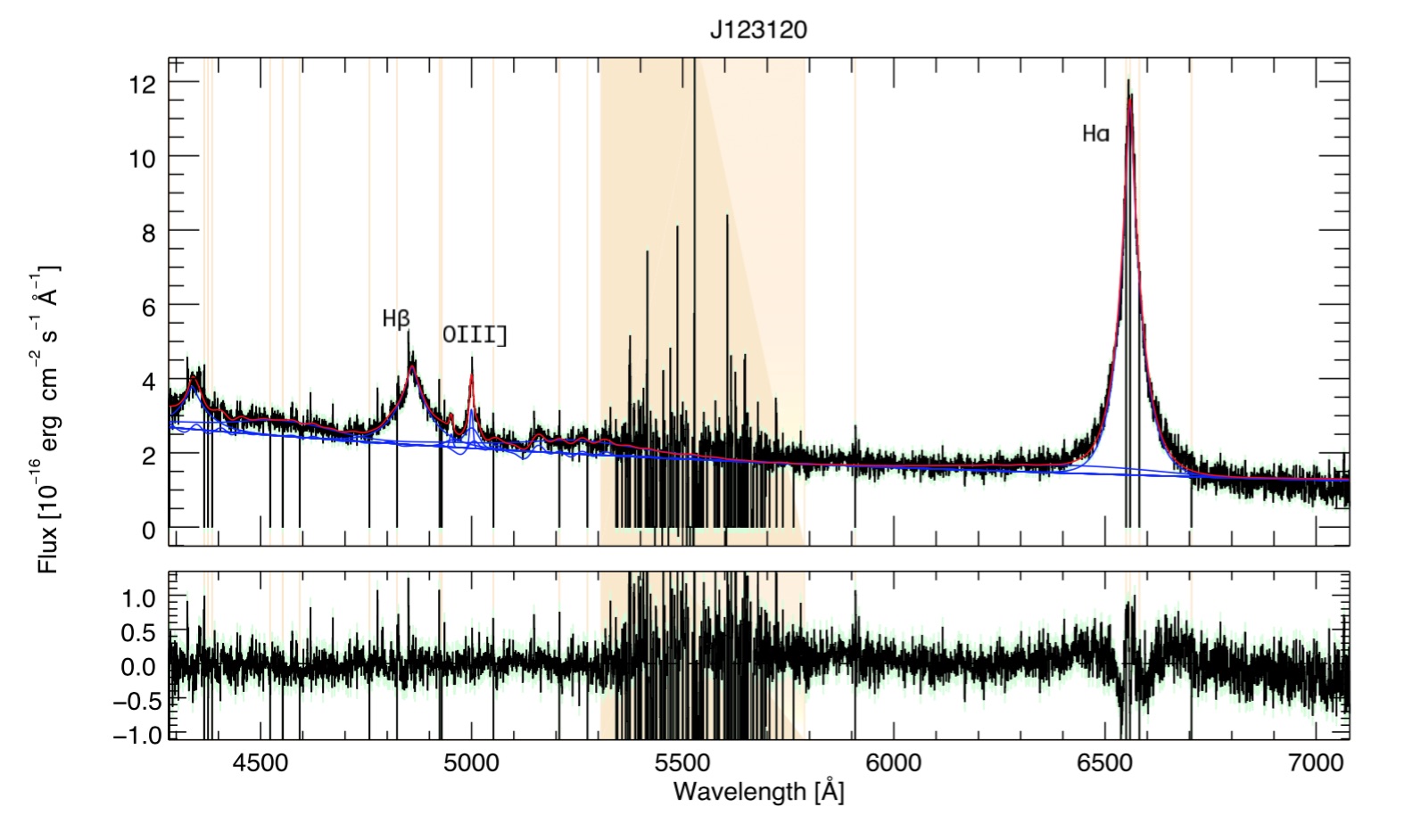}}
   {\includegraphics[scale=0.165]{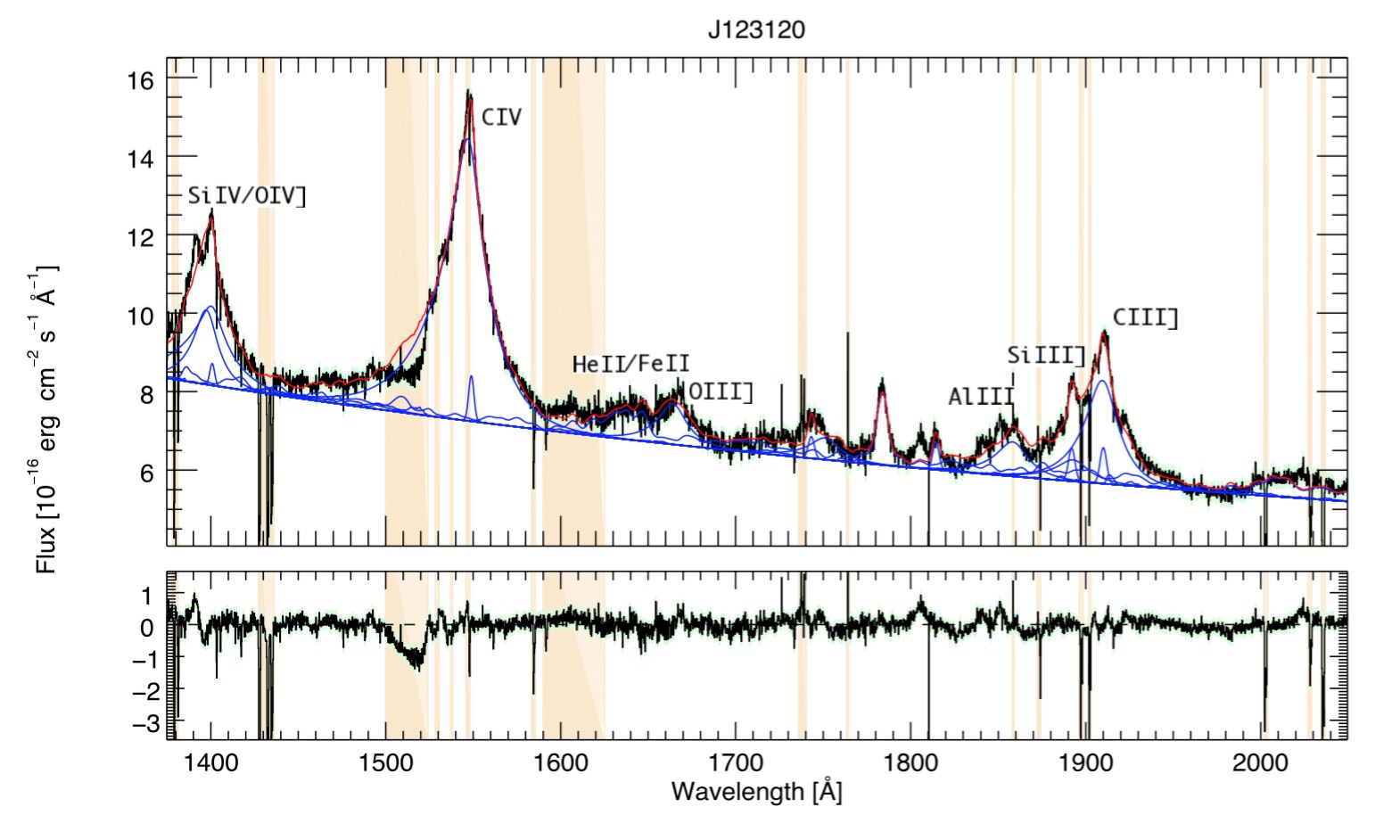}}
   {\includegraphics[scale=0.165]{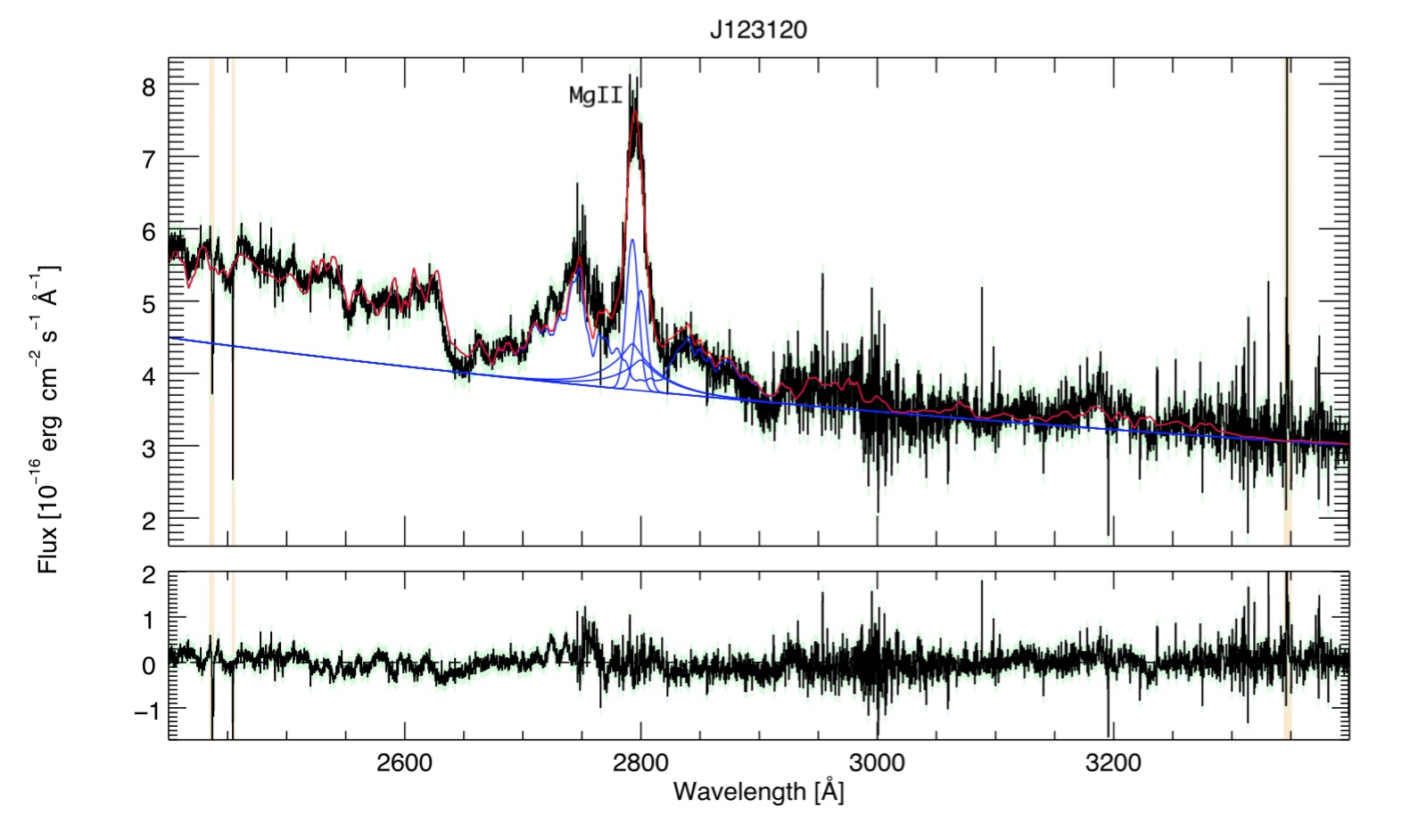}}
   {\includegraphics[scale=0.26]{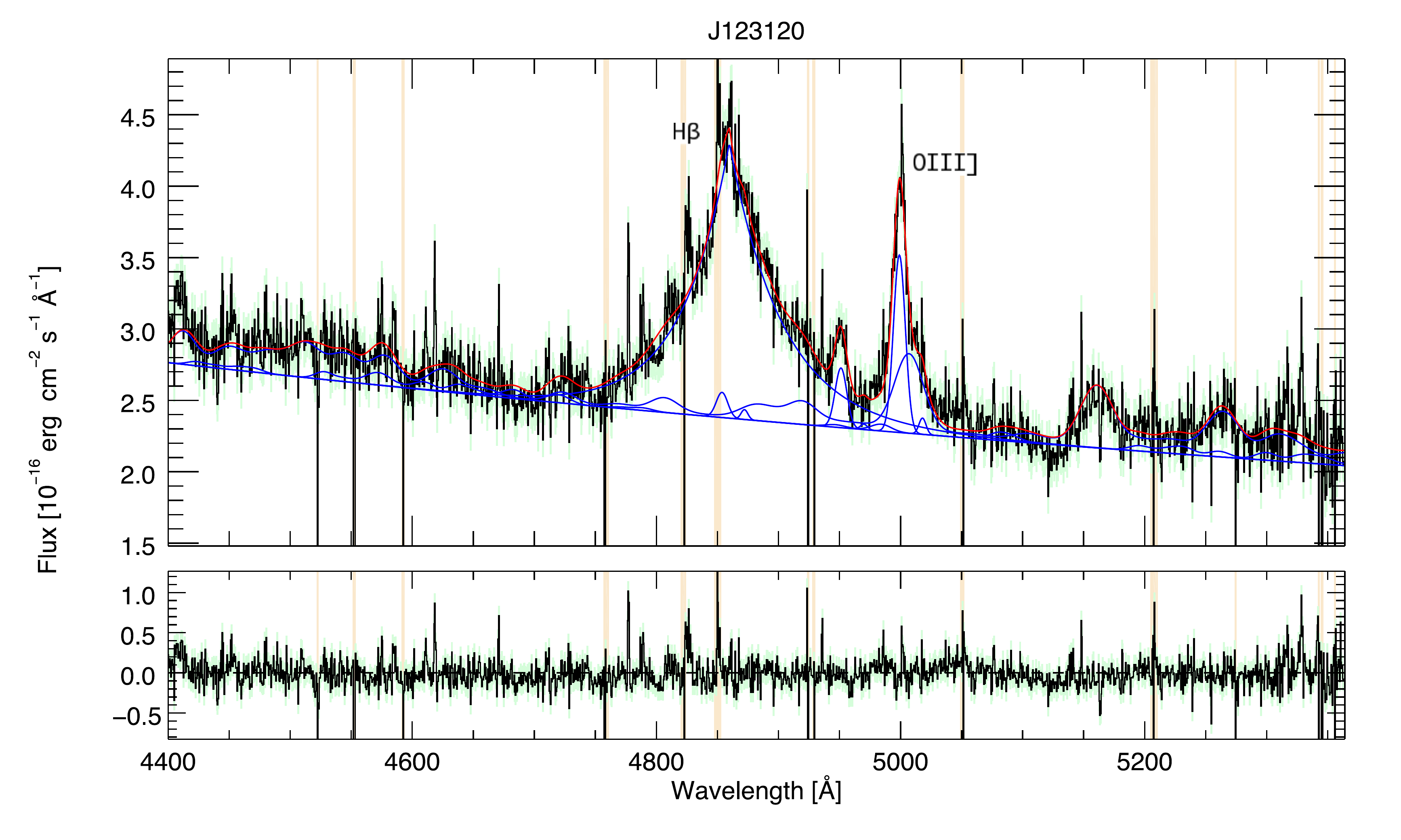}}
   {\includegraphics[scale=0.26]{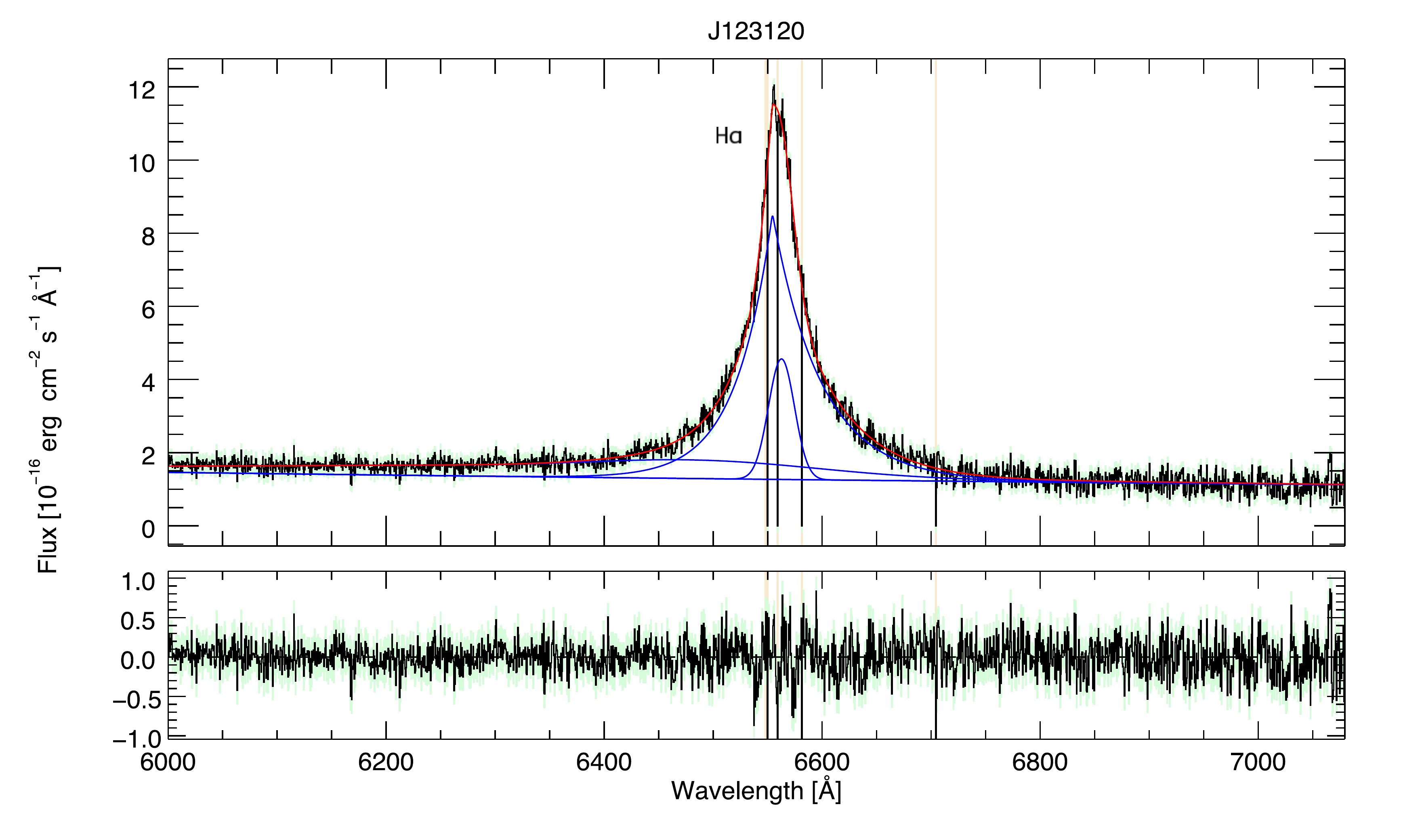}}
 \caption{Same as in the figure above but for J123120. For this source we consider also the presence of narrow components.}
 \label{fig:fits_figures_J123120}
\end{figure*}
\begin{figure*}
   {\includegraphics[scale=0.165]{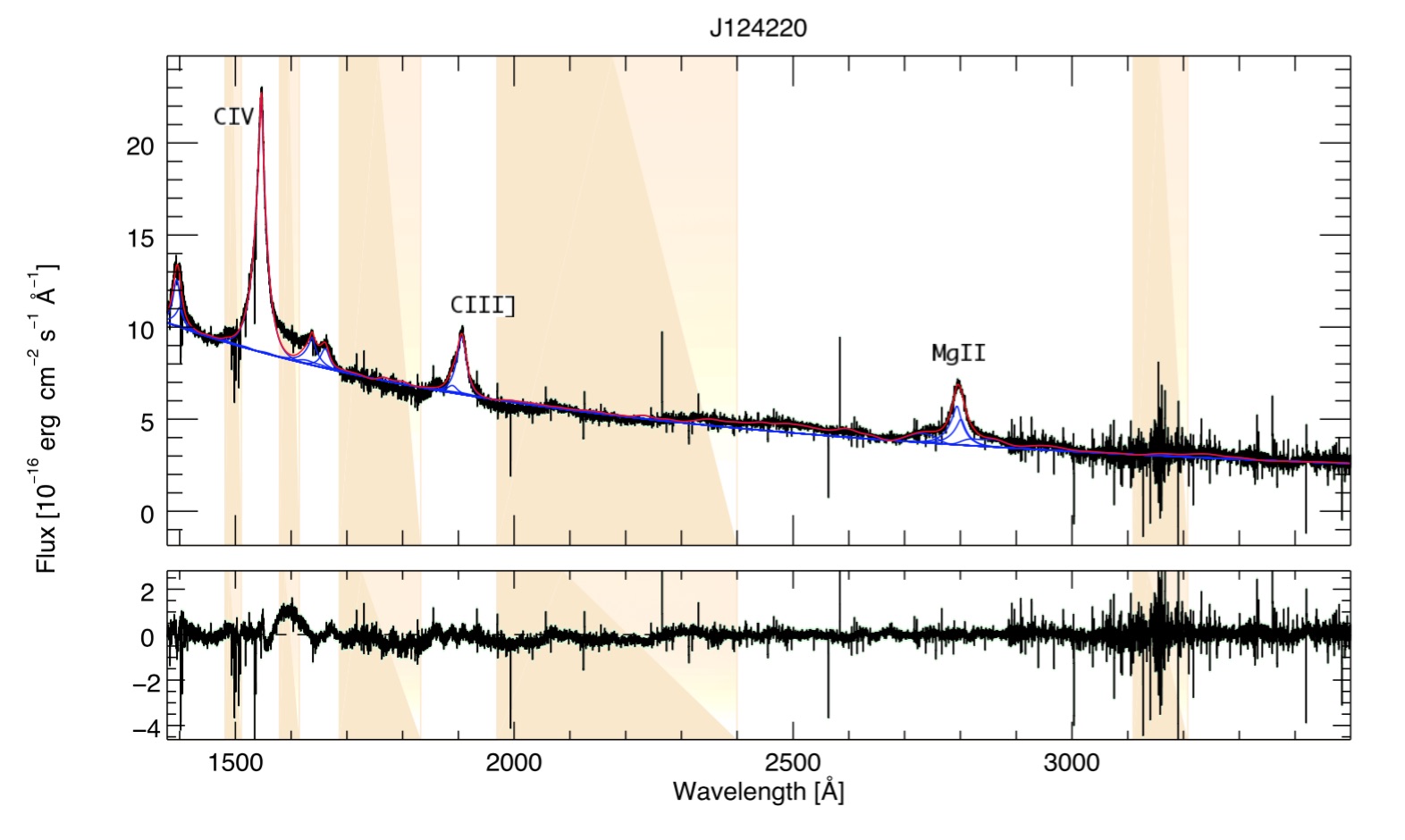}}
   {\includegraphics[scale=0.165]{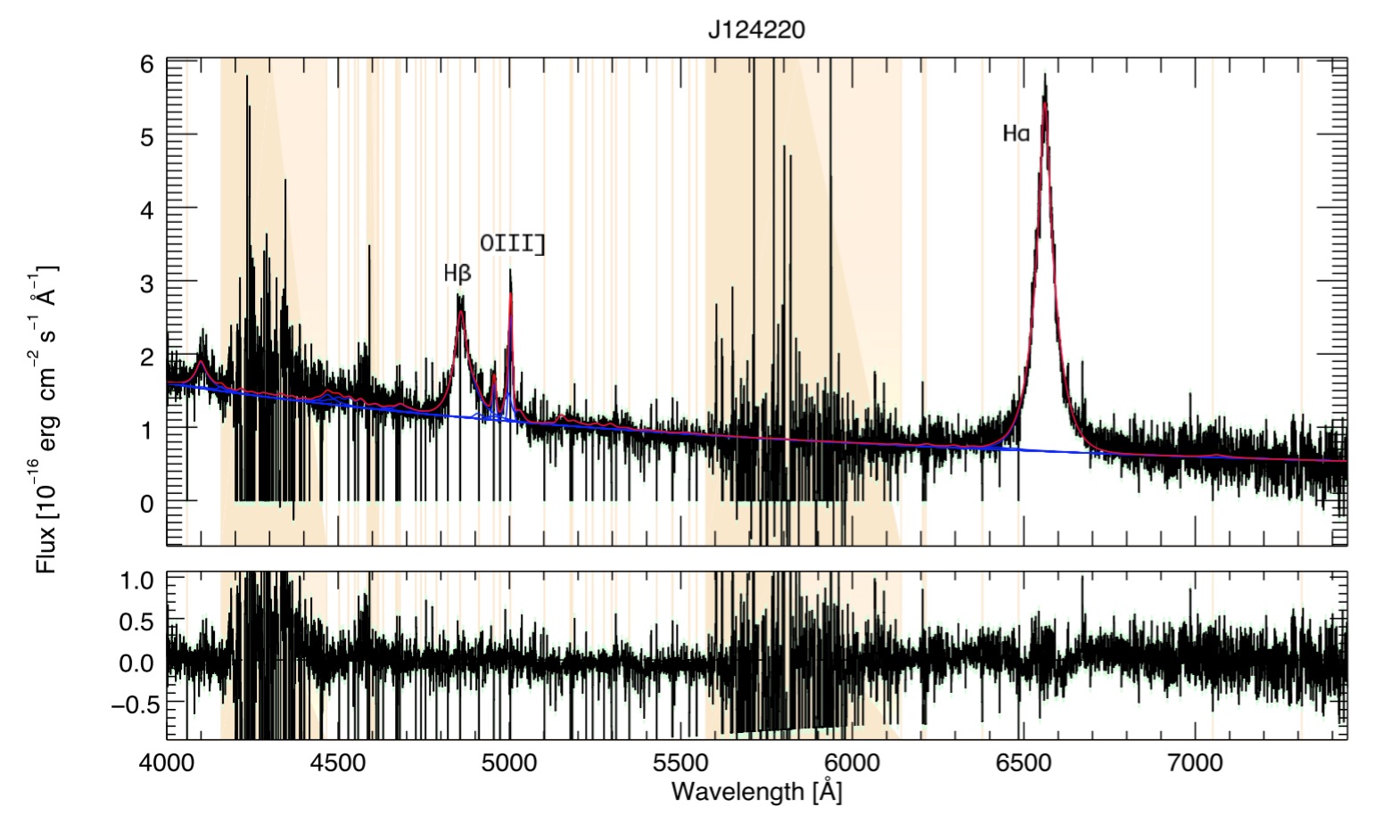}}
   {\includegraphics[scale=0.165]{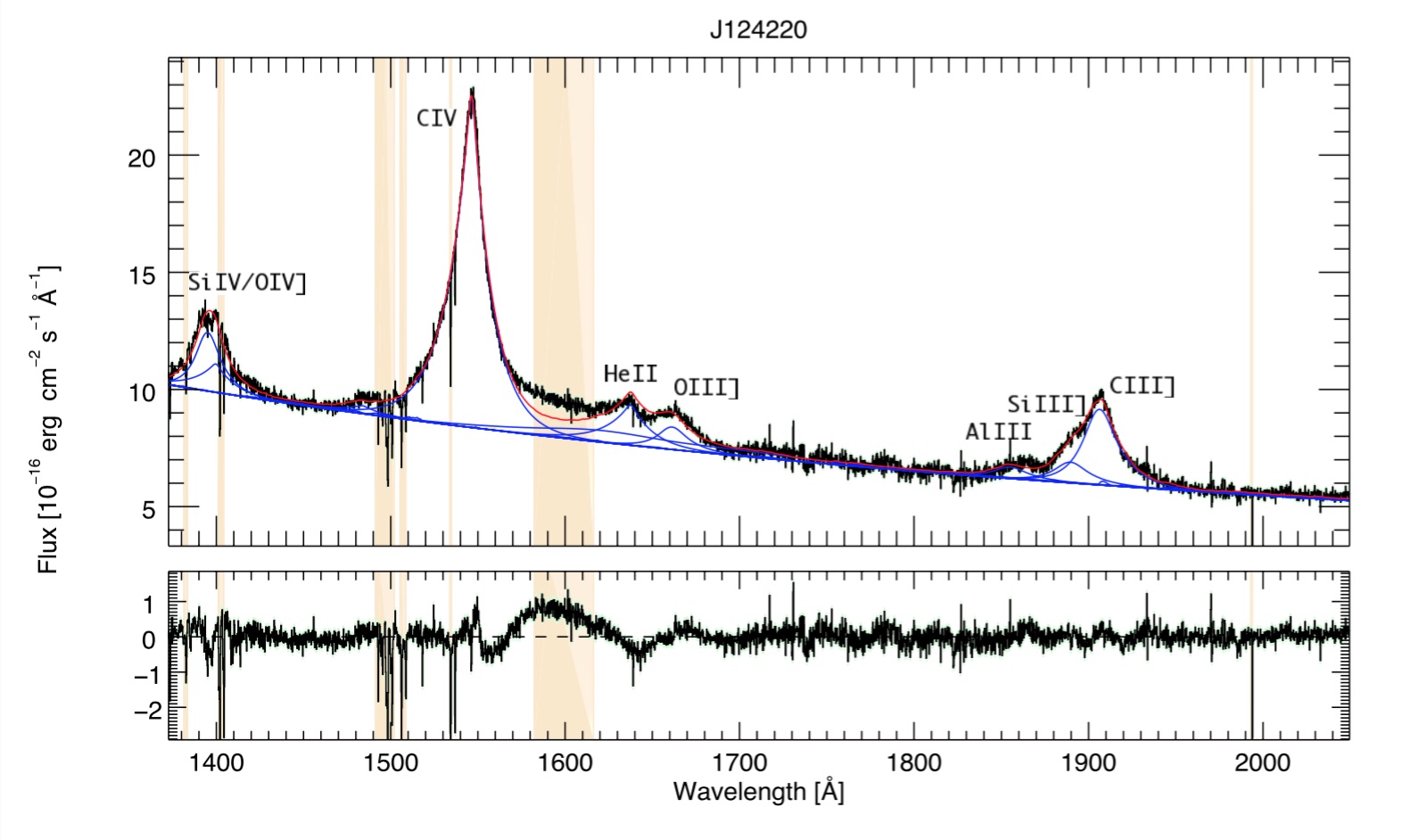}}
   {\includegraphics[scale=0.165]{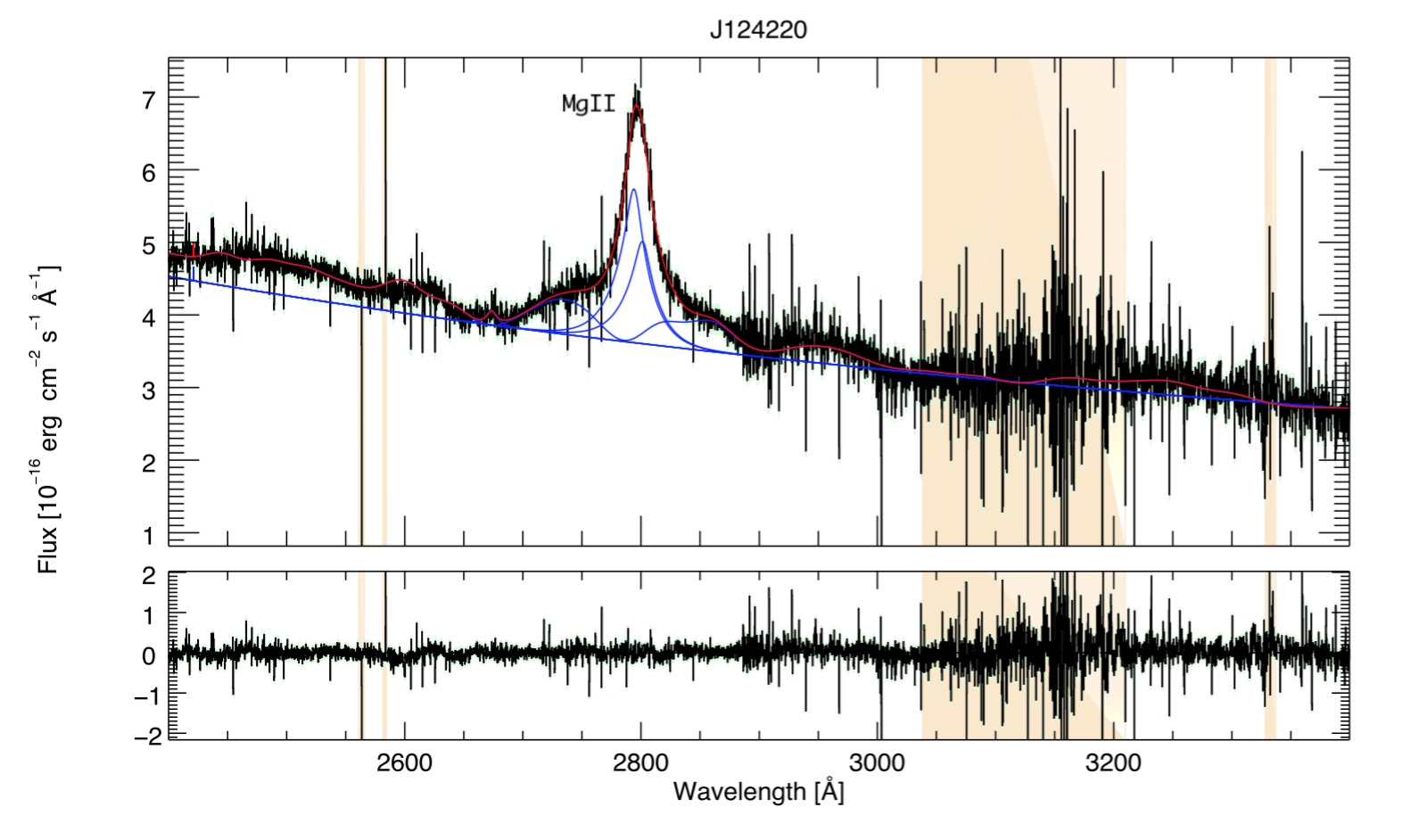}}
   {\includegraphics[scale=0.26]{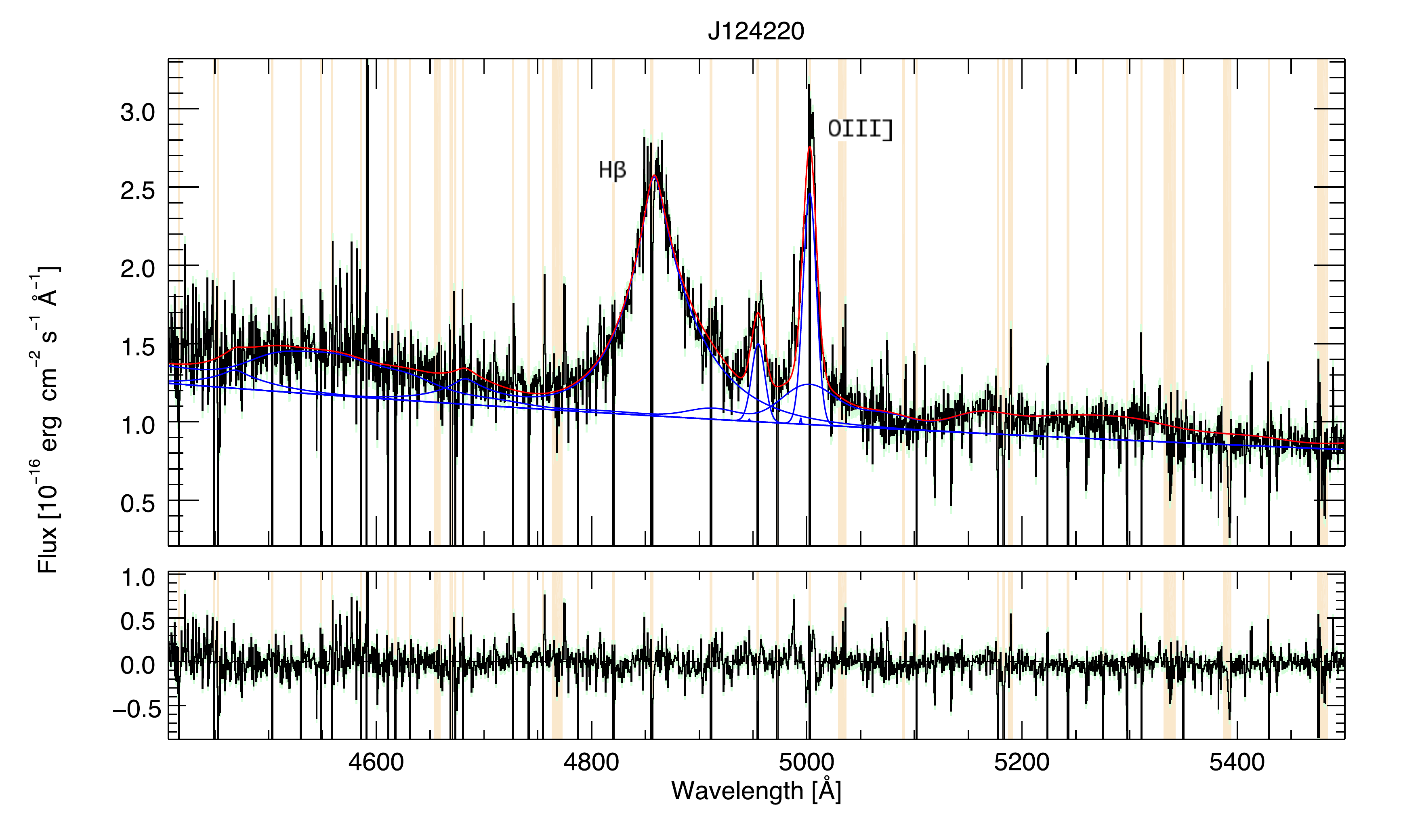}}
   {\includegraphics[scale=0.26]{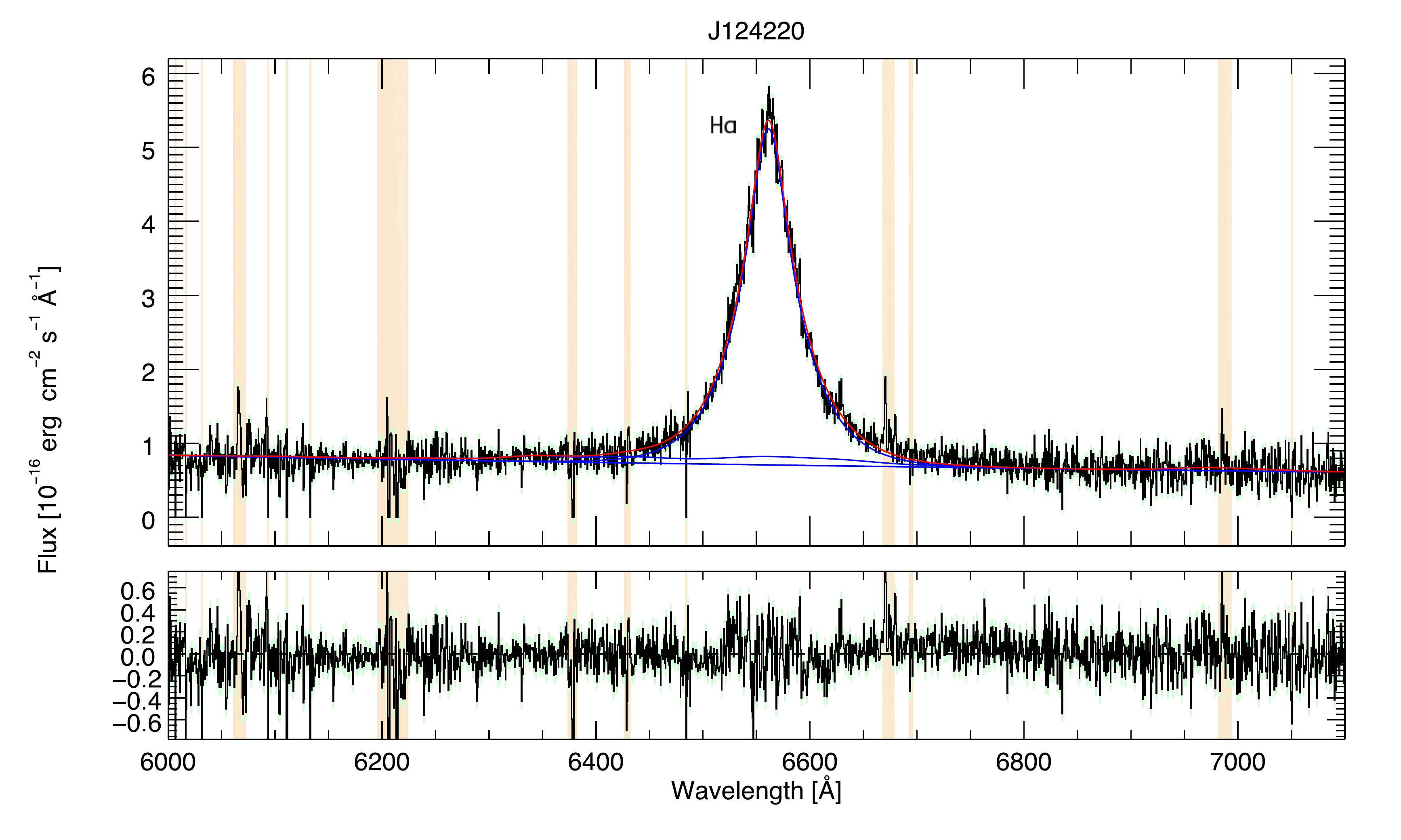}}
 \caption{Same as in the figure above but for J124220.}
 \label{fig:fits_figures_J124220}
\end{figure*}
\end{appendix}

\end{document}